\documentclass[conference]{IEEEtran}

\usepackage{xcolor} 
\usepackage[noend]{algpseudocode}
\usepackage{array}
\usepackage{amsfonts}
\usepackage{graphicx}
\usepackage{bm,mathrsfs}
\usepackage{epstopdf}
\usepackage{mathtools}
\usepackage{tikz,pgfplots}
\usetikzlibrary{snakes,arrows,shapes,trees}
\usepackage{amsmath,amsthm}
\usepackage{amsopn}
\usepackage{listings}
\usepackage{adjustbox}
\usepackage{longtable}
\usepackage{multirow}
\usepackage{hyperref}
\usepackage{pgfplots}
\usepackage{pgfplotstable}
\usepackage{grffile}
\usetikzlibrary{arrows,shapes,plotmarks}
\usepgfplotslibrary{groupplots}
\usetikzlibrary{matrix}
\usepackage{siunitx}
\usepackage[switch]{lineno}
\usepackage{enumitem}

\usepackage{titlesec}
\newcommand{\cref}[2]{\hyperref[#2]{#1~\ref*{#2}}}
\newcommand{\figref}[1]{\hyperref[#1]{Fig.~\ref*{#1}}}
\newcommand{\secref}[1]{\hyperref[#1]{Sec.~\ref*{#1}}}
\newcommand{\tabref}[1]{\hyperref[#1]{Tab.~\ref*{#1}}}
\newcommand{\eqnref}[1]{\hyperref[#1]{Eq.~(\ref*{#1})}}
\newcommand{\Algref}[1]{\hyperref[#1]{Algorithm~\ref*{#1}}}
\newcommand{\pd}[2]{\frac{\partial #1}{\partial #2}} 
 
\hypersetup{
	colorlinks,
	linkcolor={blue!50!black},
	citecolor={blue!50!black},
	urlcolor={blue!80!black},
	anchorcolor = {blue!80!black},
	filecolor = {blue!80!black},
	menucolor = {blue!80!black},
	runcolor = {blue!80!black}
}

\pgfplotsset{compat=1.8}
\usepackage{soul}
\usepackage{rotating}
\usepackage{url}
\usepackage{algorithm}

\usepackage{enumitem}
\usepackage{subcaption}
\usepackage[square,sort,comma,numbers]{natbib}

\newcommand{\mvec}{\textsc{matvec}}

\algnewcommand{\LeftComment}[1]{\Statex \(\triangleright\) #1}

\newcommand{\Stampede}{\href{https://www.tacc.utexas.edu/systems/stampede2}{Stampede2}}
\newcommand{\Frontera}{\href{https://frontera-portal.tacc.utexas.edu/}{Frontera}}
\newcommand{\petsc}{\href{https://www.mcs.anl.gov/petsc/}{PETSc}}
\newcommand{\paraview}{\href{https://www.paraview.org/
}{ParaView}}

\pgfplotsset{
compat=1.8,
legend image code/.code={
\draw[mark repeat=2,mark phase=2]
plot coordinates {
(0cm,0cm)
(0.15cm,0cm)        
(0.3cm,0cm)         
};%
}
}

\definecolor{cpu3}{HTML}{F44336}
\definecolor{cpu4}{HTML}{2196F3}
\definecolor{cpu1}{HTML}{4CAF50}
\definecolor{cpu2}{HTML}{FFC107}
\definecolor{gpu3}{HTML}{EF9A9A}
\definecolor{gpu4}{HTML}{90CAF9}
\definecolor{gpu1}{HTML}{A5D6A7}
\definecolor{gpu2}{HTML}{FFE082}

\definecolor{cpu5}{HTML}{9932CC}

\definecolor{sq_b1}{RGB}{37,52,148}
\definecolor{sq_b2}{RGB}{44,127,184}
\definecolor{sq_b3}{RGB}{65,182,196}
\definecolor{sq_b4}{RGB}{127,205,187}
\definecolor{sq_b5}{RGB}{199,233,180}
\definecolor{sq_b6}{RGB}{255,255,204}

\definecolor{sq_r1}{RGB}{189,0,38}
\definecolor{sq_r2}{RGB}{240,59,32}
\definecolor{sq_r3}{RGB}{253,141,60}
\definecolor{sq_r4}{RGB}{254,178,76}
\definecolor{sq_r5}{RGB}{254,217,118}
\definecolor{sq_r6}{RGB}{255,255,178}

\definecolor{sq_g1}{RGB}{0,104,55}
\definecolor{sq_g2}{RGB}{49,163,84}
\definecolor{sq_g3}{RGB}{120,198,121}
\definecolor{sq_g4}{RGB}{173,221,142}
\definecolor{sq_g5}{RGB}{217,240,163}
\definecolor{sq_g6}{RGB}{255,255,204}

\definecolor{div_c1}{RGB}{230,171,2}
\definecolor{div_c2}{RGB}{102,166,30}
\definecolor{div_c3}{RGB}{231,41,138}
\definecolor{div_c4}{RGB}{117,112,179}
\definecolor{div_c5}{RGB}{217,95,2}
\definecolor{div_c6}{RGB}{27,158,119}
\definecolor{div_c7}{RGB}{215,48,39}

\definecolor{div_d1}{RGB}{215,25,28}
\definecolor{div_d2}{RGB}{253,174,97}
\definecolor{div_d3}{RGB}{255,255,191}
\definecolor{div_d4}{RGB}{171,217,233}
\definecolor{div_d5}{RGB}{44,123,182}
\definecolor{ao}{RGB}{0.0, 128, 0.0}
\usepackage{dirtytalk}

\newcommand{\added}[1]{\textcolor{black}{#1}}
\usepackage[normalem]{ulem}

\begin{document}
\title{Scalable adaptive algorithms for next-generation multiphase flow simulations
}
\author{\IEEEauthorblockN{Kumar Saurabh\textsuperscript{\textsection}}
\IEEEauthorblockA{\textit{Department of Mechanical Engineering,} \\
\textit{Iowa State University, USA}\\
\href{mailto:maksbh@iastate.edu}{\texttt{maksbh@iastate.edu}}}\\

\IEEEauthorblockN{Hari Sundar}
\IEEEauthorblockA{\textit{School of Computing,} \\
\textit{University of Utah, USA}\\
\href{mailto:hari@cs.utah.edu}{\texttt{hari@cs.utah.edu}}}

\and
\IEEEauthorblockN{Masado Ishii\textsuperscript{\textsection}}
\IEEEauthorblockA{\textit{School of Computing,} \\
\textit{University of Utah, USA}\\
\href{mailto:masado@cs.utah.edu}{\texttt{masado@cs.utah.edu}}}\\
\and
\IEEEauthorblockN{Makrand A. Khanwale}
\IEEEauthorblockA{\textit{Center for Turbulence Research,} \\
\textit{Stanford University, USA}\\
\href{mailto:khanwale@stanford.edu}{\texttt{khanwale@stanford.edu}}}\\
\IEEEauthorblockN{Baskar Ganapathysubramanian}
\IEEEauthorblockA{\textit{Department of Mechanical Engineering,} \\
\textit{Iowa State University, USA}\\
\href{mailto:baskarg@iastate.edu}{\texttt{baskarg@iastate.edu}}}
}

\maketitle
\begingroup\renewcommand\thefootnote{\textsection}
\footnotetext{Equal contribution}
\endgroup
\begin{abstract}
High-fidelity flow simulations are indispensable when analyzing systems exhibiting multiphase flow phenomena. The accuracy of multiphase flow simulations is strongly contingent upon the finest mesh resolution used to represent the fluid-fluid interfaces. However, the increased resolution comes at a higher computational cost. In this work, we propose algorithmic advances that aim to reduce the computational cost without compromising on the physics by selectively detecting key regions of interest (droplets/filaments) that require significantly higher resolution. 
The framework uses an adaptive octree--based meshing framework that is integrated with ~\petsc{}'s linear algebra solvers. We demonstrate scaling of the framework up to 114,688 processes on TACC's ~\Frontera{}. Finally, we deploy the framework to simulate one of the most resolved simulations of primary jet atomization. This simulation -- \textit{equivalent} to 35 trillion grid points on a uniform grid -- is 64$\times$ larger than current state--of--the--art simulations and provides unprecedented insights into an important flow physics problem with a diverse array of engineering applications.  


\end{abstract}


\section{Introduction}
Multiphase flows -- more specifically, two-phase flows, where one fluid (water, paint, melts, etc.) interacts with another fluid (usually air) -- are ubiquitous in natural and engineered systems. Examples include natural phenomena from breaking waves and cloud formation to engineering applications like printing, additive manufacturing, and all types of spraying operations in healthcare and agriculture.

High-fidelity modeling of two-phase flows has been an indispensable strategy for understanding, designing, and controlling such phenomena. For instance, insights from high-fidelity modeling of the fluid-fluid interface interactions have produced accurate, low-cost coarse-scale models used to simulate large systems like chemical/biological reactors~\citep{joshi2015computational}. High-fidelity simulations also serve as the basis for the optimization-based design of micro-scale systems, with applications in bio-microfluidics and advanced manufacturing.

\begin{figure}
    \centering
    \includegraphics[width=\linewidth]{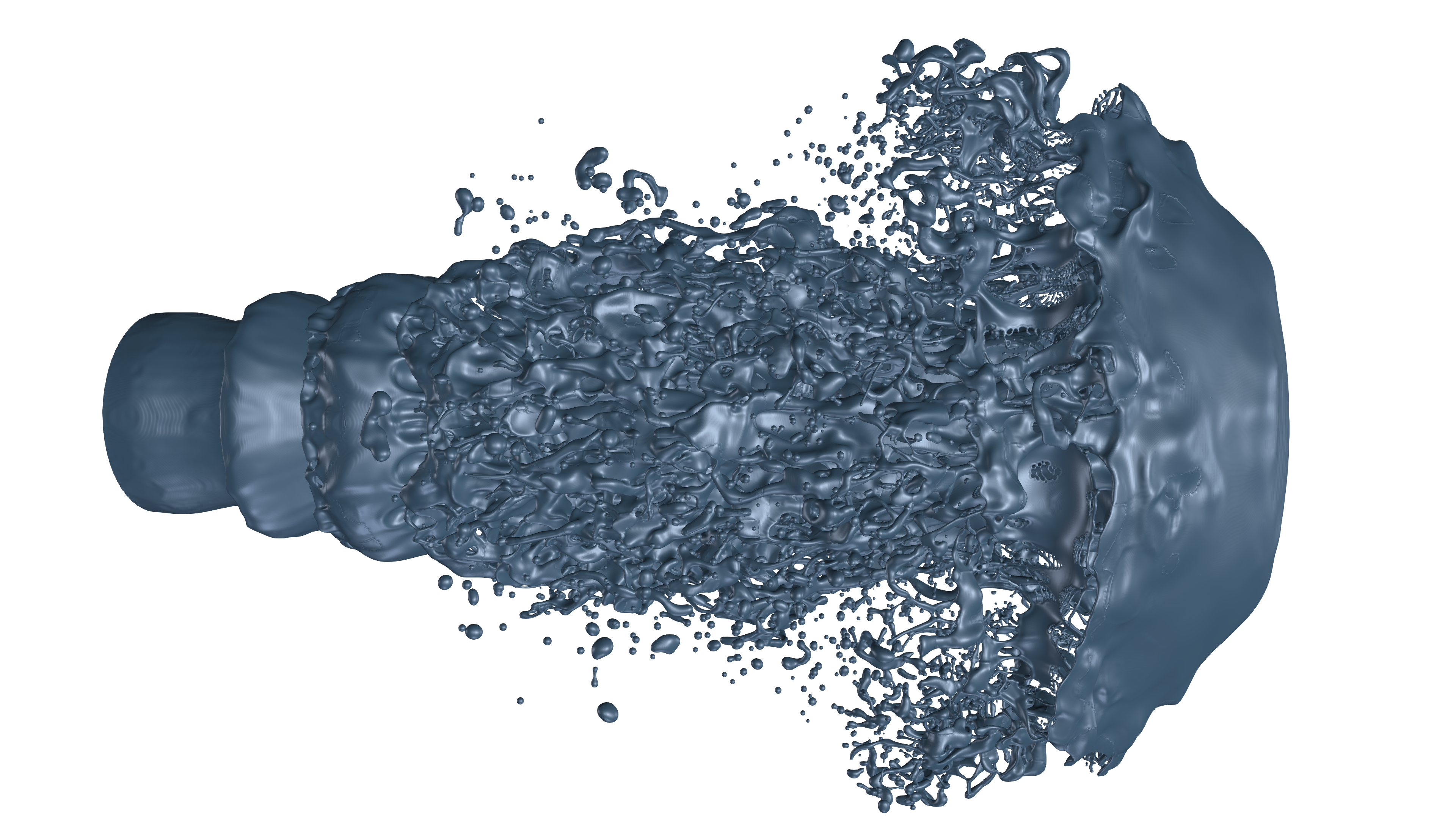}
    \caption{Snapshot of a multiphase flow simulation of primary jet atomization of liquid diesel jet at $T  = \SI{6.3}{\micro\second}$. 
    This simulation contains $\sim 3B$ unknowns and is solved using $\sim 200,000$ node hours on TACC \Frontera.
    }
    \label{fig:jet}
\end{figure}

High-fidelity modeling, specifically through interface-resolving simulations of two-phase flows, is difficult due to the wide range of spatial and temporal scales, especially under turbulent conditions (\figref{fig:jet}). Such approaches typically require spatially adaptive and temporally higher-order methods to capture the relevant phenomena. Additionally, the size of these simulations requires the development of scalable algorithms. This remains a very active area of research~\citep{Herrmann2010} with significant space for improvement. We briefly describe the landscape of interface-resolving simulations and identify a critical challenge that this paper resolves. 

Interface-resolved two-phase modeling can be broadly divided into two main categories -- sharp--interface methods and diffuse--interface methods. The sharp interface methods rely on representing the interface with a sharp, discontinuous function (for instance, the volume of fluids (VOF)), whereas the diffuse interface methods smear out this sharp interface to construct a diffuse, continuous representation of the interface (e.g., conservative diffuse interface, Cahn--Hillard Navier--Stokes (CHNS)). 
Sharp interface methods have been the state--of--the--art for high-fidelity simulations of turbulent multiphase flows ~\cite{fuster2009simulation,tomar2010multiscale,
pairetti2020mesh,herrmann2010detailed}. 
These methods involve solving a PDE for transporting the discontinuous volume fraction function.  An interface reconstruction procedure is generally required to construct normals. 
The sharp interface methods suffer from artificial breakup -- also known as numerical surface tension, when the interfacial features (droplets/filaments) are comparable in length scale $r$ to the grid size $\Delta x$ ~\cite{gorokhovski2008modeling}. There has been some progress in the development of computational techniques that reduces this effect, with good examples being \textit{interface reconstruction} techniques~\citep{lu2018direct,chiodi2020advancement}, \textit{moment-of-fluid} methods~\cite{jemison2015filament}, and \textit{manifold death} algorithms~\cite{chirco2022manifold}. These methods, however, remain computationally complex. 

Diffuse interface methods, particularly CHNS, avoid any interface reconstruction process. However, diffuse interface methods---especially under low mesh resolution---also suffer from artificial breakup and mass loss, but for completely different reasons. When the thickness of the diffuse interface, $\epsilon$, becomes comparable to the length scale, $r$, of the flow features of interest ($\epsilon/r \sim \mathcal{O}(1)$), it leads to bound violations of the interface tracking variable and artificial breakup. Additionally, these smaller fluid structures can disappear via absorption into nearby larger structures (due to coarsening of Cahn-Hilliard), affecting summary statistics like the number of droplets. 

This brief review of state-of-art is primarily to make the case that both sharp and diffuse interface approaches suffer from issues of artificial breakup and numerical artifacts emanating from insufficient mesh resolution. Adaptive meshing approaches are an elegant strategy to ensure that the local mesh size is always smaller than the \textit{local length scales of interest}, thus allowing interface tracking approaches to reliably capture multi-scale features in a computationally efficient manner. However, automatically identifying these local regions of interest is itself non-trivial. This serves as the primary motivation for the current work, and we make the following contributions:
\begin{enumerate}[label=\roman*, wide]
    \item We develop scalable algorithms to identify the spatial regions of interest in the computational domain where the flow features become comparable to the mesh resolution, i.e., regions where $\epsilon/r  \approx \mathcal{O}(1)$. This is especially needed in phenomena exhibiting droplets and fluid filaments, where such targeted resolution is critical for performing cost-effective simulations
    \item We improve on existing octree refinement and coarsening algorithms to accelerate remeshing and decrease the associated overhead, especially for multi-level refinements. This is essential for simulations where the element sizes drop substantially. For instance, in the canonical example of primary jet atomization shown in the results section, element sizes vary by three orders of magnitude to accurately resolve fluid features varying by nine orders of magnitude in volume. This contrasts with existing approaches, where refinement or coarsening of the octrees is done level by level. \cite{saurabh2021industrial,saurabh2021scalable,fernando2018massively,popinet2003gerris,bangerth2012algorithms}
    \item We analyze the data movement during matrix and vector assembly of the finite element method (FEM) operators used and propose a new data layout for local matrix and vector assembly that is more amenable to avoiding cache misses.
    \item We showcase the proposed algorithm to simulate a canonical problem of primary jet atomization and produce one of the highest resolution datasets of this phenomenon (see Fig.~\ref{fig:jet}). Our simulation required over 200,000 node hours on TACC Frontera, is equivalent to $35$ trillion grid points on a uniform mesh and is $64\times$ more resolved than the current state--of--the--art simulations. 
\end{enumerate}
  
\section{Methods}
Our algorithmic developments apply to all interface-capturing methods. We specifically implement and illustrate these developments on a thermodynamically consistent Cahn--Hillard Navier--Stokes (CHNS) model, which is a diffuse interface approach.
\subsection{Numerical methods}
\label{sec:LocalCahn}
\noindent\textbf{\textit{Governing equations:}}
Let $\phi$ be the phase field variable that tracks the location of the phases and varies
smoothly between +1 to -1 with a characteristic diffuse interface thickness $\epsilon$, and $v_i$ be the $i^{th}$ component of the mixture velocity of the two phases. $\eta_+$ and $\rho_+$ ($\eta_-$ and \added{$\rho_-$}) represents the viscosity and density of the +1 phase fluid (-1 phase fluid). The thermodynamically consistent CHNS equations~\citep{Khanwale2023projection,guo2017mass,shen2015decoupled,anderson1998diffuse,feng2006fully} are written as follows:
\begin{enumerate}[label=\roman*]
    \item Momentum equation:
    \footnotesize
    \begin{equation*}
    \begin{split}
         &\pd{\left(\rho(\phi) v_i\right)}{t} + \pd{\left(\rho(\phi)v_iv_j\right)}{x_j} + \frac{1}{Pe}\pd{\left(J_jv_i\right)}{x_j}+\frac{Cn}{We} \pd{}{x_j}\left({\pd{\phi}{x_i}\pd{\phi}{x_j}}\right)
         \\
         &+\frac{1}{We}\pd{p}{x_i} - \frac{1}{Re}\pd{}{x_j}\left({\eta(\phi)\pd{v_i}{x_j}}\right) - \frac{\rho(\phi)\hat{{g_i}}}{Fr} = 0,\\
         & \quad\quad \text{where,} \quad J_i = \frac{\left(\rho_- - \rho_+\right)}{2\;\rho_+ Cn} \, m(\phi)\pd{\mu}{x_i},
    \end{split}
    \end{equation*}
    \normalsize
    \item Solenoidality and Continuity:
    \footnotesize
    \begin{equation*}
        \pd{v_i}{x_i} = 0, \quad \pd{\rho(\phi)}{t} + \pd{\left(\rho(\phi)v_i\right)}{x_i}+
	\frac{1}{Pe} \pd{J_i}{x_i} = 0,
    \end{equation*}
    \normalsize
    \item Cahn--Hillard Equations
    \footnotesize
    \begin{equation*}
    \begin{split}
        &\pd{\phi}{t} + \pd{\left(v_i \phi\right)}{x_i} - \frac{1}{PeCn} \pd{}{x_i}\left(m(\phi){\pd{\mu}{x_i}}\right) = 0 \\
        &\mu = \psi'(\phi) - Cn^2 \pd{}{x_i}\left({\pd{\phi}{x_i}}\right)
    \end{split}
    \end{equation*}
    \normalsize
\end{enumerate}

where $\rho(\phi)$ is the non--dimensional mixture density given by $\left({\rho_+ - \;\rho_-}/{2\rho_+}\right) \phi + \left({\rho_+ + \;\rho_-}/{2\rho_+}\right)$, the non--dimensional mixture viscosity $\eta\left(\phi\right)$ is given by $\left({\eta_+ - \;\eta_-}/{2\eta_+}\right) \phi + \left({\eta_+ + \;\eta_-}/{2\eta_+}\right)$. 
\normalsize
We use a degenerate mobility $m(\phi)$ given by $\sqrt{1-\phi^2}$.  
Non-dimensional parameters are as follows: Peclet, $Pe = \frac{u_{r} L_{r}^2}{m_{r}\sigma}$; Reynolds, $Re = \frac{u_{r} L_{r}}{\nu_{r}}$; Weber, $We = \frac{\rho_{r}u_{r}^2 L_{r}}{\sigma}$; Cahn, $Cn = \frac{\varepsilon}{L_{r}}$; and Froude, $Fr = \frac{u_{r}^2}{gL_{r}}$, with $u_{r}$ and $L_r$ denoting the reference velocity and length, respectively.
\noindent\textbf{\textit{Numerical discretization:}}
We utilize the two-block projection scheme~\citep{Khanwale2023projection} that is based on a projection-based pressure splitting of Navier--Stokes (NS). Overall, the scheme involves four solutions per block:
\begin{enumerate}[label=\roman*]
    \item \textbf{CH-Solve}: As the first step, we solve the advective Cahn--Hillard equation. This is solved as a non--linear solver in a fully implicit manner.
    \item \textbf{NS-Solve}: 
    We compute mixture density and viscosity as a function of $\phi$ 
    from the CH-Solve (from step 1) and use it to solve the momentum equations. 
    We use a semi-implicit discretization based on the Crank-Nicolson scheme that results in linearized momentum equations. 
    \item \textbf{PP-Solve}: 
    The projection-based splitting results in a variable density pressure Poisson equation to compute the pressure. 
    \item \textbf{VU-solve}: As the last step for each block, we correct the velocity predicted in the NS-solve to get the solenoidal velocity.  
\end{enumerate}
    To minimize the memory footprint, we split the final VU-solve to update the velocity once per each velocity component. This reduces the assembled matrix size from $N\times \mathrm{DIM}$ to $N$, where $N$ is the number of independent DOF, and $\mathrm{DIM}$ is the spatial dimension (2D/3D). We note that the mass matrix in the VU-solve did not need to be (re)computed and assembled for each of the $\mathrm{DIM}$ separately and is reused as long as the mesh does not change. This also reduces the preconditioner computation cost, as well as the associated communication costs.

     We utilize a second-order semi-implicit formulation (as used in \cite{Khanwale2023projection}) that exhibits a milder time-stepping restriction $\Delta t$ compared to fully explicit counterparts. The explicit parts of the discretization allow for the linearization of NS, which avoids an expensive setup of Newton iteration. This represents a good balance between time-step restriction and computational complexity. 
   
    All simulations are performed using linear basis functions (spatially second-order convergence). The approach is extensible to arbitrary order basis functions with Continuous Galerkin (CG) elements. 

\noindent\textbf{\textit{Local Cahn:}}
Diffuse interface methods, including CHNS,  under low mesh resolution suffer from artificial breakup and mass loss. Particularly for CHNS, this limitation arises when the interface thickness $\epsilon$ has a comparable length scale to the length scale of fluid features (droplets/filaments/sheets), leading to an artificial breakup. We non-dimensionalize the interface thickness, $\epsilon$, with the system characteristic length to get the Cahn number, $Cn$. 
A smaller $Cn$ represents a thinner diffuse interface which allows the capture of more accurate physics. A smaller $Cn$ compared to the other interface features would prevent an artificial breakup. 
However, a decrease in $Cn$ everywhere comes at an additional cost of increased mesh resolution. A rough estimate indicates that a $2\times$ decrease in $Cn$ leads to approximately $8\times$ increase in mesh resolution. In this work, we develop novel scalable algorithms to tackle this issue, where we selectively decide the regions where the interface thickness has become comparable to the length scale of interfacial features and reduce $Cn$ only in those regions. We refer to this approach as ``local Cahn". This is in contrast with the current state--of--the--art approach for CHNS simulation, which relies on a constant Cahn number throughout the domain.
This approach balances the accurate capture of relevant physics and the controlled increase in mesh resolution, which is critical to tackling engineering problems. 

\subsection{Identification of key features}
\label{sec:FeatueDetection}

Multiphase flow simulations produce flow structures exhibiting small scales. Examples include fluid sheets (that exhibit small length scales in one dimension), fluid filaments (exhibiting small length scales in two dimensions), and fluid droplets (exhibiting small length scales in all three dimensions). Identifying these flow features is the first step to accurately resolving them.
We now present algorithms to detect such regions of interest.  The Cahn number in these local regions of interest is decreased. 
We first highlight the salient features of the algorithm on a uniform mesh before discussing the parallel deployment on octree meshes.

\subsubsection{Uniform mesh case:}
\label{sec:UniformDilation}
Our algorithm is inspired by the classic image-processing idea of erosion and dilation. The basic idea is that small structures like drops and filaments will disappear with erosion followed by dilation, whereas larger structures will not be affected. ~\figref{fig:schematic} briefly illustrates the algorithmic idea via two examples: identifying a small drop (\figref{fig:Drop}) and identifying the elongated filament connecting two larger fluid masses (\figref{fig:filament}). 

\input{tikz/erosionDilaiton}

 We perform a series of following steps to achieve the desired goal:
\begin{enumerate}[label=\roman*]
    \item $\phi$, the phase field variable, continuously varies from -1 to 1. As a first step, we convert this field (or image) to a binary 0/1 representation $\phi_{BW}$ by thresholding at a prescribed value, $\delta$: (denoted by $\mathcal{T}(\phi)$ in ~\figref{fig:schematic}).
    
    \footnotesize
    \begin{equation*}
\phi_{BW} = \begin{cases}
1 & \quad \phi \leq  \delta, \\
0 & \quad \phi > \delta.
\end{cases}
\end{equation*}
\normalsize
We choose $\delta$ (usually $\pm0.8$) such that the bulk phase is 0 and immersed phase is 1. 
\item Second, we repeatedly apply a morphological erosion operator to shrink the region of immersed ($\phi_{BW} = 1$) phase. (denoted by $\mathcal{E}(\phi)$ in ~\figref{fig:schematic}).
\item Third, we repeatedly apply a morphological dilation operator to expand the region\footnote{We perform more steps of dilation compared to erosion} of $\phi_{BW}$ (denoted by $\mathcal{D}(\phi)$ in ~\figref{fig:schematic}). 
\item Finally, we compare the dilated image ($\phi_{BWd}$) with the original thresholded image $\phi_{BW}$ (denoted by $\mathcal{S}(\phi)$) to identify the regions of interest. Essentially, these regions are those that appear as 1 in  $\phi_{BW}$ while appearing as 0 on $\phi_{BWd}$. 
\item After these regions are identified, we decrease the Cahn number and refine the mesh. 
\end{enumerate} 

\subsubsection{Challenges with octree mesh:}
The above steps work well on a uniform mesh with accessible neighbors; however, performing these steps in a parallel distributed setting on adaptive meshes poses the following challenges:
\begin{itemize}
    \item The octree-based mesh has elements with non-uniform sizes.
    \item CG-based finite element (FEM) data structures may not have neighbor information. This is a particularly challenging issue with unstructured meshes, as elements can have a varying number of neighbors with no plausible upper limits.
    \item Hanging nodes involve the interpolation of values from parent to child elements. So, $\phi_{BW}$  will not have only two distinct values: 0 and 1, but can have arbitrary values between 0 and 1, depending on the type of hanging nodes - edge hanging or face hanging in 3D. 
\end{itemize}

\subsubsection{Parallel deployment on octree meshes}
We extend the idea of the approach illustrated in ~\secref{sec:UniformDilation} to octree meshes in parallel settings, albeit with some modifications.

\begin{itemize}
    \item The first modification involves changing the limits of $\phi_{BW}$. We modify $\phi_{BW}$ as:
    \footnotesize
    \begin{equation}
        \phi_{BWO} = \begin{cases}
        1 & \quad \phi \leq  \delta, \\
        -1 & \quad \phi > \delta.
    \end{cases}
    \label{eq:thresholdOctree}
    \end{equation}
    \normalsize
    The above modification in the definition of $\phi_{BWO}$ is purely a mathematical convenience in detecting the interface elements and helps in subsequent erosion and dilation algorithms.
    
    \item \textbf{Interface elements:} An element contains interface when:
    \footnotesize
    \begin{equation}
        \lvert \sum_{i=1}^{nodes} \phi_{BWO} \rvert \neq nodes
        \label{eq:interface}
    \end{equation}
    \normalsize

    \item \textbf{Erosion:} Erosion step involves visiting the interfacial element (\eqnref{eq:interface}) and making all the nodal values to be -1.
    \item \textbf{Dilation:} Dilation step involves the reverse of the erosion step. It involves visiting the interface element and making all the nodal values +1.
\end{itemize}
    
We now define the key steps associated with identifying the features of interest in octree-based meshes. \Algref{alg:localCahnIdentifier} lists the major steps involved in identifying the local regions of interest and locally decreasing the $Cn$ number. We begin by converting the continuous phase field variable $\phi$ to $\phi_{BW,o}$. We then impose erosion and dilation as a series of ~\mvec{} operations 
\footnote{\mvec{} operations are at the heart of FEM computations and has been shown to have excellent scaling ~\cite{ishii2019solving,saurabh2021scalable}. Each ~\mvec{} operation involves a single pass over all the local elements with associated ghost exchange. We note that these ghost exchange communication is typically overlapped with computation ~\cite{ishii2019solving,saurabh2021scalable,fernando2017machine,burstedde2011p4est}.} 
(\Algref{alg:erodeDilate}).  To perform each erosion step, we make a single pass over the elements in a distributed fashion, with each processor looping over its own unique set of elements. When the element containing the interface is visited, the erosion step is triggered, which involves converting all the nodal values of that element to -1. As previously mentioned, the octree mesh contains elements at various levels of refinement. To address this challenge, we define a base level $b_l$, typically the finest level in the octree mesh.
Additionally, we maintain a counter on how often the element has been visited for erosion. An erosion is triggered when the counter equals $(b_l - l)$, where $l$ is the current level of the octree mesh. For example, an element two times coarser than the finest mesh, i.e., $b_l - l = 2$, will have to wait for two iterations before erosion can finally be triggered. This gives a way to balance the difference in levels of octree mesh. Similarly, as before, we then perform dilation steps to follow the erosion steps. The dilation is also executed similarly to account for the difference in octree levels. We perform a larger number of dilation steps compared to the erosion steps to compensate for the thresholding encountered during $\phi_{BWO}$, thus ensuring that no region that is a part of a larger region gets marked for high refinement, leading to a drastic increase in the problem size. Finally, we make a last pass over the elements and mark it with reduced Cahn based on the following condition:

\footnotesize
\begin{equation}
    \sum_{i=1}^{nodes}\phi_{BWO}   = nodes   \quad \& \quad 
    -\sum_{i=1}^{nodes}\phi_{BW,d} = nodes 
\label{eq:condnCahn}
\end{equation}
\normalsize
~\eqnref{eq:condnCahn} states that we need to reduce Cahn for the element that has all the nodes marked as +1 after thresholding and -1 after extra dilation. This is analogous to the view in the uniform mesh, except we are processing element-wise rather than node-wise. 
(\Algref{alg:elemCahn})

Our approach depends on the number of erosion and dilation steps - a parameter that controls the identified regions. To prevent very small regions of local $Cn$, we further perform erosion and dilation on the elemental $Cn$ vector 
(\Algref{alg:dilateCahn}). 
This is done by first converting the elemental $Cn$ vector to the nodal $Cn$ vector and performing the subsequent erosion and dilation steps on the nodal vectors. This serves a two-fold purpose. First, it removes the local island of small $Cn$ that can hinder the solver convergence. Secondly, the extra dilation steps help to pad the surrounding regions. This relaxes the need to perform region detection every timestep.
Once we have identified the elements for reducing $Cn$, we refine the regions of the interface marked by $|\phi|<\delta$.

\begin{algorithm}[b!]
  \scriptsize
    \caption{\textsc{LocalCahnIdentifier:} \footnotesize{Identify regions of Local Cahn}}
    \label{alg:localCahnIdentifier}
    \begin{algorithmic}[1]
\Require Phase field vector $\phi$, Reference level ($b_l)$
\Ensure Elemental Local $Cn$
\item[]
\State $\phi_{BWO}$   $\leftarrow $ \textsc{Threshold}($\phi_{BW}$) \Comment{~\eqnref{eq:thresholdOctree}}
\State $\phi_{BW,e} \leftarrow $ \textsc{ErodeDilate}($\phi_{BWO}$, \textsc{Erosion},numErodeSteps,$b_l$) \Comment{\Algref{alg:erodeDilate}}
\State $\phi_{BW,d} \leftarrow $ \textsc{ErodeDilate}($\phi_{BW,e}$, \textsc{Dilate},numDilateSteps,$b_l$)\Comment{\Algref{alg:erodeDilate}}
\State elemental\_Cahn $\leftarrow$ \textsc{ElementalCahn}($\phi_{BWO}$,$\phi_{BW,d}$,Cahn1,Cahn2)\Comment{\Algref{alg:elemCahn}}
\item[]
\Return elemental\_$Cn$
    \end{algorithmic}
\end{algorithm}

\begin{algorithm}[b!]
  \scriptsize
    \caption{\textsc{ErodeDilate:} \footnotesize{Erosion and dilation step}}
    \label{alg:erodeDilate}
    \begin{algorithmic}[1]
\Require Nodal vector $\phi_{BWO}$, Stage (\textsc{Erosion/Dilation})
\Ensure Nodal vector $\phi_{BW,e}$ or $\phi_{BW,d} $, depending on the stage
\item[]
\State vec  $\leftarrow \phi_{BWO}$ 
\If{Stage == \textsc{Erosion}}
\State val $\leftarrow$ -1 \Comment{Value to set for Erosion}
\Else
\State val $\leftarrow$ +1 \Comment{Value to set for Dilation}
\EndIf
\State vec $\leftarrow \phi_{BWO}$ 
\For{i $\leftarrow$ 0:num\_steps}
\State vec\_ghosted $\leftarrow$ GhostRead (vec)
\State vec\_temp $\leftarrow$ vec\_ghosted \Comment{Temporary vector}
\State counter  $\leftarrow$ [] 
\For{elem $\leftarrow$ local\_elems}
\State phi\_elemental $\leftarrow$  vec\_ghosted[elem] \Comment{Copy the nodal values}
\State phi\_Sum $\leftarrow$ 0
\For{node $\in$ num\_local\_nodes}
\State phi\_Sum += phi\_elemental[node]
\EndFor
\State has\_Interface $\leftarrow$ (abs(phi\_sum) $\neq$ num\_local\_nodes)
\If{has\_Interface}
\State $l_c \leftarrow$ level[elem]
\If{counter[elem] $== l_c - b_l$ } \Comment{Check for level}
\For{node $\leftarrow$ num\_local\_nodes}
\State vec\_temp[elem][node] $\leftarrow$ val
\EndFor
\EndIf
\EndIf
\EndFor
\State  vec\_ghosted $\leftarrow$ vec\_temp
\State  vec $\leftarrow$ GhostWrite(vec\_ghosted)
\EndFor
\item[]
\Return vec
    \end{algorithmic}
\end{algorithm}

\begin{algorithm}[b!]
  \scriptsize
    \caption{\textsc{ElementalCahn:} \footnotesize{Identify the element with local $Cn$}}
    \label{alg:elemCahn}
    \begin{algorithmic}[1]
\Require Binary phase field vector $\phi_{BWO}$, Phase field vector after dilation $\phi_{BW,d}$, $Cn$ values ($Cn_1$, $Cn_2$) : ($Cn_1$ $<$ $Cn_2$)
\Ensure Elemental Local $Cn$
\item[]
\State elemental\_Cahn $\leftarrow$ []
\State vec\_ghosted\_o $\leftarrow$ GhostRead( $\phi_{BWO}$)
\State vec\_ghosted\_d $\leftarrow$ GhostRead( $\phi_{BW,d}$)
\For{elem $\leftarrow$ local\_elems}
\State phi\_elemental\_o $\leftarrow$  vec\_ghosted\_o[elem] \Comment{Copy the nodal values}
\State phi\_elemental\_d $\leftarrow$  vec\_ghosted\_d[elem] \Comment{Copy the nodal values}
\State $\phi_o$ $\leftarrow$ 0
\State $\phi_d$ $\leftarrow$ 0
\For{node $\in$ num\_local\_nodes}
\State $\phi_o$ += phi\_ghosted\_o[node]
\State $\phi_d$ += phi\_ghosted\_d[node]
\EndFor
\If{$\phi_o$==num\_local\_nodes \& $\phi_d$== \textbf{-} num\_local\_nodes } 
\State elemental\_$Cn$[elem] = $Cn_2$
 \Else
\State elemental\_$Cn$[elem] = $Cn_1$
\EndIf
\EndFor
\item[]
\Return elemental\_$Cn$
    \end{algorithmic}
\end{algorithm}

\begin{algorithm}[b!]
  \scriptsize
    \caption{\textsc{ErodeDilateCahn:} \footnotesize{Expand $Cn$ regions and remove islands}}
    \label{alg:dilateCahn}
    \begin{algorithmic}[1]
\Require elemental\_$Cn$, Reference level ($b_l)$
\Ensure Elemental Local $Cn$
\item[]
\State $Cn_{nodal} \leftarrow $ []
\State $Cn_{nodal} \leftarrow $ +1
\For{elem $\in$ local\_elems}
\If{elemental\_$Cn$ == $Cn_2$}
\For{nodes $\in$ num\_local\_nodes}
\State $Cn_{nodal}[node] \leftarrow -1$
\EndFor
\EndIf
\EndFor
\State  $Cn_{nodal}$ $\leftarrow$ \textsc{ErodeDilate}( $Cn_{nodal}$,\textsc{Erosion},numStageErosion,$b_l$) \Comment{\Algref{alg:erodeDilate}}
\State  $Cn_{nodal}$ $\leftarrow$ \textsc{ErodeDilate}( $Cn_{nodal}$,\textsc{Dilation},numStageDilation,$b_l$) \Comment{\Algref{alg:erodeDilate}}
\For{elem $\in$ local\_elems}
\State $Cn_{elem} \leftarrow $ $Cn_{nodal}$[elem]
\If{\texttt{anyof}($Cn_{elem}$) == -1}
\State elemental\_$Cn$ $\leftarrow$ $Cn_1$
\Else
\State elemental\_$Cn$ $\leftarrow$ $Cn_2$
\EndIf
\EndFor
\item[]
\Return elemental\_$Cn$
    \end{algorithmic}
\end{algorithm}

The overall complexity of the proposed approach still scales as $\mathcal{O}(N)$ per dilation and erosion step, the same as the uniform case. Although we only demonstrate the algorithm deployment for octree-based meshes, this can be extended for any other class of unstructured meshes. The algorithm proposed relies on efficient ~\mvec{} operations, which also form the core kernel for FEM operation. Thus, it can be applied to any unstructured FEM code without major changes to the framework. Furthermore, the same algorithm can be extended to detect multiple levels of $Cn$ - each with its own set of numbers of erosion and dilation steps. 
\subsection{Refinement \& Intergrid transfer}

\subsubsection{Adaptive refinement algorithms}
Octrees, specifically 2:1 linearized octrees, have been successfully deployed for various scientific simulations. However, the past literature has mostly focused on refining or coarsening the octree only by a single level. This can lead to significant overhead during remeshing, especially when the coarsest to the finest mesh differs by several levels. \footnote{In our specific problem of interest, the coarsest to the finest mesh varies by 11 levels. \secref{sec:Atomization}} This section focuses on the algorithmic development needed to circumvent the restriction of one-level refinement and coarsening, thereby enabling multi-level refinement.




\paragraph{Refinement in serial}
Refinement, in the context of octrees, amounts to substituting sets of leaves by their descendants. \Algref{alg:multi-level-refine} takes a sorted leaf set and replaces each leaf with its descendants at the specified level in sorted order. 
The traversal begins at the root. If the current input octant needs refining below the current traversal depth, only then are child subtrees
traversed; otherwise omitted. After
processing the descendant subtree of the current octant, the input
pointers are advanced. The next subtree is traversed only if it overlaps
with the next input octant.

\begin{algorithm}[b!]
  \scriptsize
    \caption{\textsc{Refine:} \footnotesize{Replace leaf octants by sorted descendants}}
    \label{alg:multi-level-refine}
    \begin{algorithmic}[1]
\Require List of leaf octants and desired levels, with pointers $\textrm{oct}_\textrm{in}$ and  $\textrm{level}_\textrm{in}$. %
         Mutable list to capture output octants, $\textrm{oct}_\textrm{out}$. %
         Root of the subtree to traverse, $R$, and its SFC orientation, $\textrm{sfc}$.
\Ensure $\textrm{oct}_\textrm{out}$ contains the descendants of %
        $\textrm{oct}_\textrm{in}$ at $\textrm{level}_\textrm{in}$.
\item[]
\If{not $R$ overlaps $\textrm{oct}_\textrm{in}[0]$}
  \Return
\EndIf
\If{$\textrm{level}(R) < \textrm{level}_\textrm{in}[0]$}
  \For{$0 \leq c < 2^\textrm{dim}$}
    \State refine($\textrm{oct}_\textrm{in}$, $\textrm{level}_\textrm{in}$, $\textrm{oct}_\textrm{out}$, $\textrm{child}(R, \textrm{sfc}, c)$, $\textrm{curve}(\textrm{sfc}, c)$)
  \EndFor
\Else
  \State push($\textrm{oct}_\textrm{out}, R$)
\EndIf
\While{$\textrm{oct}_\textrm{in} < \textrm{end}$ and $\textrm{oct}_\textrm{in} = R$}
  \State ++$\textrm{oct}_\textrm{in}$
  \State ++$\textrm{level}_\textrm{in}$
\EndWhile
    \end{algorithmic}
\end{algorithm}

\paragraph{Refinement in parallel}
Obtaining a refined set of octants can be done independently on each process, in an embarrassingly parallel way. However, once the refinement is completed, the 2:1-balance condition must be restored after the refinement. 
In the context of incomplete octrees \cite{saurabh2021scalable}, refinement near the domain boundaries can create void descendants, either directly or resultant of 2:1 balancing. In such cases, void descendants of boundary-intercepted octants need to be discarded.

\paragraph{Coarsening in serial}

Analogous to refinement, the octree can be coarsened by various levels throughout the domain.
An integer is attributed to the coarsest acceptable level to which each octant can be promoted.
However, unlike refinement, the decision to coarsen a subtree depends on the consensus of all descendants.
Consensus is defined based on two requirements:
An ancestor $A$ of an input leaf is output if (i) no descendants of $A$ in the input require $A$ to be refined and if (ii) the same cannot be said of the parent of $A$.

\Algref{alg:multi-level-coarsen} traverses the tree while iterating in order over the input---just
as for refinement---but here, the output is pushed and popped at every subtree,
rather than the leaves only.
Per requirement (i), the inputs within the current subtree must be read before deciding
whether the root of the subtree is too coarse.
Therefore, the root is emitted in the post-order action only after checking all descendants.
On the other hand, by (ii), the root of the current subtree could be too fine,
depending on inputs outside the subtree.
Therefore, we leave it to the parent to retract emitted roots of child subtrees if needed.

\begin{algorithm}[b!]
  \scriptsize
    \caption{\textsc{Coarsen:} \footnotesize{Replace leaf octants by sorted ancestors}}
    \label{alg:multi-level-coarsen}
    \begin{algorithmic}[1]
\Require List of leaf octants and desired levels, with pointers $\textrm{oct}_\textrm{in}$ and  $\textrm{level}_\textrm{in}$. %
         Mutable list to capture output octants, $\textrm{oct}_\textrm{out}$. %
         Root of the subtree to traverse, $R$, and its SFC orientation, $\textrm{sfc}$.
\Ensure $\textrm{oct}_\textrm{out}$ contains the ancestors of %
        $\textrm{oct}_\textrm{in}$ at $\textrm{level}_\textrm{in}$, %
        subject to the least amount of coarsening voted by any descendant.
\item[]
  \State coarsen\_to $\leftarrow$ 0
  \If{not $R$ overlaps $\textrm{oct}_\textrm{in}$[0]}
    \Return
  \EndIf
  \If{level($R$) $<$ level($\textrm{oct}_\textrm{in}$[0])}
    \State pre\_size $\leftarrow$ size($\textrm{oct}_\textrm{out}$)
    \Comment{Remember where to emit parent}
    \For{$c$ in range(0,pow2(dim))}
      \State $L_c$ $\leftarrow$
          coarsen($\textrm{oct}_\textrm{in}$, $\textrm{level}_\textrm{in}$, $\textrm{oct}_\textrm{out}$, child($R$, sfc, $c$), curve(sfc, $c$))
	  \State coarsen\_to $\leftarrow$ max(coarsen\_to, $L_c$)
	\EndFor
    \If{coarsen\_to $\leq$ level($R$)}
      \State repeat(pop($\textrm{oct}_\textrm{out}$), size($\textrm{oct}_\textrm{out}$) - pre\_size)
      \Comment{Undo child emit}
      \State push($\textrm{oct}_\textrm{out}$, $R$)
      \Comment{Emit parent octant}
    \EndIf
  \Else
    \State push($\textrm{oct}_\textrm{out}$, $R$)
    \State coarsen\_to $\leftarrow$ $\textrm{level}_\textrm{in}$[0]
  \EndIf
  \While{$\textrm{oct}_\textrm{in}$.nonempty() and $\textrm{oct}_\textrm{in}$[0] = $R$}
    \State ++$\textrm{oct}_\textrm{in}$
    \State ++$\textrm{level}_\textrm{in}$
  \EndWhile
  \Return coarsen\_to
    \end{algorithmic}
\end{algorithm}



We note that our approach is similar to~\citet{burstedde2011p4est}
but with subtle differences. 
In contrast to using a Boolean callback function evaluated on every subtree as a coarsening criterion, we tailored our algorithm for an upfront declaration of maximum change in level for each octant.  
Additionally, we relax the assumption that octree needs to be complete, which is important for simulations involving complex domains.


\paragraph{Coarsening in parallel}

The coarsening, unlike refinement, requires extra communication to guarantee consensus among all the descendants.
\begin{algorithm}[b!]
  \scriptsize
    \caption{\textsc{ParCoarsen:} \footnotesize{Replace distributed leafs by partitioned and sorted ancestors}}
    \label{alg:par-coarsen}
    \begin{algorithmic}[1]
\Require List of leaf octants and desired levels, $\textrm{oct}_\textrm{in}$ and  $\textrm{level}_\textrm{in}$. %
         MPI communicator $\textrm{comm}$.
\Ensure Returns the ancestors of%
        $\textrm{oct}_\textrm{in}$ at the consensus of $\textrm{level}_\textrm{in}$, with duplicates removed.
  \item[] \LeftComment{first coarsening pass}
  \State tentative = coarsen($\textrm{oct}_\textrm{in}$,  $\textrm{level}_\textrm{in}$)
  \Comment{\Algref{alg:multi-level-coarsen}}
  \item[] \LeftComment{exchange tentative coarse octants at partition endpoints}
  \State $r \leftarrow$ rank(comm)
  \State head\_tail[1] $\leftarrow$ $\{$ tentative[0], tentative[-1] $\}$
  \State head\_tail[2] $\leftarrow$ send\_recv(head\_tail[1], to:$r-1$, from:$r+1$, comm)
  \State head\_tail[0] $\leftarrow$ send\_recv(head\_tail[1], to:$r+1$, from:$r-1$, comm)
  \item[] \LeftComment{find overlapped inputs}
  \State overlap\_prev = [],  overlap\_next = []
  \If{level(tail[0]) $\leq$ level(head[1])}
    \State overlap\_prev = [$x$ if overlaps($x$, tail[0]) for $x$ in $\textrm{oct}_\textrm{in}$]
  \EndIf
  \If{level(head[2]) $<$ level(tail[1])}
    \State overlap\_next = [$x$ if overlaps($x$, head[2]) for $x$ in $\textrm{oct}_\textrm{in}$]
  \EndIf
  \item[] \LeftComment{repartition overlapped inputs (omitted: also repartition levels)}
  \State $\textrm{oct}_\textrm{pre} \leftarrow$ send\_recv(overlap\_next, to:$r+1$, from:$r-1$, comm)
  \State $\textrm{oct}_\textrm{post} \leftarrow$ send\_recv(overlap\_prev, to:$r-1$, from:$r+1$, comm)
  \State $\textrm{oct}_\textrm{in} \leftarrow \textrm{oct}_\textrm{in}$ - overlap\_prev - overlap\_next + $\textrm{oct}_\textrm{pre}$ + $\textrm{oct}_\textrm{post}$
  \item[] \LeftComment{second coarsening pass}
  \item[]
  \Return coarsen($\textrm{oct}_\textrm{in}$,  $\textrm{level}_\textrm{in}$)
  \Comment{\Algref{alg:multi-level-coarsen}}
    \end{algorithmic}
\end{algorithm}
\Algref{alg:par-coarsen} describes the algorithm for parallel coarsening. 
The globally sorted input octree undergoes a local coarsening pass
resulting in a locally linearized incomplete tree.
The results are tentative, as the output octants achieve consensus locally on a process,
but may not represent the global consensus.
Furthermore, coarse octants might be duplicated even if the results agree across processes.
The input must be 
repartitioned before the second coarsening pass to prevent conflicts.
Note that the required partitioning resolution
for multi-level coarsening is known only after the first coarsening attempt.

After the first local coarsening pass, the next step is finding the overlap region between consecutive processes.
Ideally, the tentative coarse octants do not overlap between two consecutive processes $p_i$ and $p_{i+1}$.
If so, the inputs on processes $p_{j \leq i}$ are independent of those on processes $p_{j > i}$.
Otherwise, the worst case for two processes is for a single coarse octant
on one process to overlap one or more octants on the other process.

In the rare case that a tentative octant is so aggressively coarsened
that it overlaps multiple remote partitions, the initial coarse octant exchange
can be considered the first step of a distributed exponential search
to find the processes containing the region of overlap.
The details of this adjusted algorithm are omitted for brevity.


\subsubsection{Distributed multi-level inter-grid transfer}

After updating the resolution of the octree mesh on a new time step
(and restoring 2:1-balance),
data from the previous grid must be transferred to the resolution and partition of the new grid.

Octree-based meshing frameworks typically support inter-grid transfers only between successive levels \cite{sundar2012parallel}.
This constraint simplifies the implementation because it means overlapping cells differ in size by a factor of at most two.
In case of multi-level refinement, constructing intermediate grids one level at a time creates significant overhead.
This section describes our methods to transfer data after refining or coarsening an arbitrary number of levels.

\paragraph{Serial algorithms}

Transferring coarse-to-fine cell-centered data is the simplest case. We
traverse the octants of old and modified grids in the SFC order simultaneously. 
Once a leaf is reached in the coarse grid, all cells overlapped in the fine
grid are accessible. The value in the coarse cell is imposed on all the
overlapped fine cells.

Coarse to fine nodal interpolation begins similarly.
Octants of both grids (coarse \& fine) are traversed together. Then, at a coarse leaf,
nodal values on all overlapped fine cells can be interpolated from the
coarse cell nodes. 
Using mesh-free techniques~\citep{ishii2019solving}, the coordinates of the relevant nodes are available in a contiguous list at this point in the traversal. The transfer is done by evaluating the field in the coarse element at each of the fine nodes. If neighboring subtrees redundantly interpolate a node, the results are consistent in each instance.

Fine-to-coarse transfers follow essentially the same structure, except
mirrored. As multiple fine unknowns are aggregated into fewer coarse
unknowns at a coarse leaf, the exact aggregation function depends on the
differential operator and the type of transfer. Cell-centered values
might be averaged. Node-centered values might be simply injected or
otherwise weighted by the transpose of the interpolation operator.

The advantage of using mesh-free grid traversals here is that the nodes
need not be initially arranged in any particular order.
The only requirement is that nodal values are tagged by their unique location code key.
The keys make it possible to route the appropriate nodes to each subtree during the traversal.
This property is needed by our distributed memory approach
of sending nodes---with their keys---to interested processes in a detached manner.

\paragraph{Parallel algorithm}
We extend inter-grid transfer to a distributed memory algorithm
structured in four steps:
i) search for grid-grid overlaps in the splitter table;
ii) detach and send coarse element nodes to the fine grid;
iii) execute serial inter-grid interpolation or aggregation;
iv) if coarsening, receive coarse element data from the fine grid and attach it.

It is possible to communicate either before or after the local inter-grid transfer,
\added{depending on whether the source grid or target grid has a coarser resolution.
We communicate \textit{before} locally upsampling coarse-to-fine,
and we communicate \textit{after} locally downsampling fine-to-coarse.}
Effectively, the total message size is reduced by choosing to communicate only the coarse data.
Also, since the number of fine elements dominates the local intergrid transfers,
this strategy uses the fine partition to keep the computational workload balanced.

\subsubsection{Scalability}
\label{sec:remesh-improvements}
In this section, we report results from the scalability tests that were essential for scaling beyond 30K processes.

\paragraph{Sorting octree keys (Allreduce and Alltoallv)}

Sorting octree keys in distributed memory is a building block of our meshing routines, such as for repartitioning, 2:1 balancing, and enumeration of nonhanging nodes. AlltoAll communication generally suffers at a large number of processors. 
To circumvent this, we adopt a hierarchical \(k\)-way staged communication pattern, similar
to other hypercube exchange sorting algorithms
\citep{sundar2013hyksort, fernando2017machine}. In the \(k\)-way
hierarchical scheme, the number of (super)partitions is kept below a
constant \(k\) for each of \(\mathcal{O}(\log_k(p))\) stages. (By
default \(k=128\), so at most, three stages are required for up to 2M processes).
Dissociating the partitions from process count reduces the storage cost of
splitter selection from \(\mathcal{O}(p)\) to \(\mathcal{O}(k)\), and
data transfer of Allreduce from \(\mathcal{O}(p)\) to
\(\mathcal{O}(k \log_k(p))\). Performing the Alltoall exchange in stages
is also a standard defense against network congestion.

\paragraph{Hierarchical communication (Comm\_split)}

During the setup phase of the \(k\)-way hierarchical exchange, the
global communicator is subdivided recursively with
\texttt{MPI\_Comm\_split}, such that the final stage involves \(k\) or
fewer processes. Splitting a communicator is a costly global operation.
Doing it repeatedly poses a serious threat to scalability. Since the
arguments to \texttt{MPI\_Comm\_split} are not dependent on the data to
be sorted, we memorize the sequence of communicators in an \emph{MPI user
cache attribute} attached to the root communicator. On later executions
of distributed octree sort, the saved sequence of communicators is
recalled without extra splitting.

\paragraph{Remote processing (Sparse Alltoallv)}

Our nodal enumeration algorithm contains an outsourcing pattern, whereby
candidate nodes are sorted globally, checked for duplication and
hangingness on remote processes, and sent back to the owners of the
originating elements. 

The last step of the algorithm uses a concept of \emph{return address}
to return nodes to originating processes. Due to the high-locality
heuristics of SFC sorted orders, the number of destination processes is
actually sparse. At the same time, the return addresses are treated as
arbitrary. Our first implementation used a raw \texttt{MPI\_Alltoall}
collective to obtain the receive counts at the destinations.
\texttt{MPI\_Alltoall} tends to produce heavy network congestion and
requires time to populate the send count array, which is linear in
process count. We noticed that this step of the algorithm had low
overhead up to 28K cores, at which point scaling halted and the overhead
blew up 15x from 28K to 56K cores. To fix this we adopted the NBX sparse
exchange algorithm \citep{hoefler2010scalable}. \added{NBX stands for non-blocking consensus algorithm for dynamic sparse data exchange which uses
non-blocking collectives and point-to-point messages to eliminate any
\(\Omega(p)\) primitives like \texttt{MPI\_Alltoall}}.
\added{\section{Software Implementation}}
\subsection{Matrix \& Vector assembly}

In this work, we interface our framework with \petsc{} data--structure to perform linear algebra operations. Specifically, we store the matrix in the form of block storage \texttt{MATMPIBAIJ}. This format has been demonstrated to be more efficient than the non-block version \texttt{MATMPIAIJ} for multi--DOF systems. The block size, here, refers to the number of DOFs per node. 

This storage format, although ideal for the global format, is not suitable for local assembly. This is primarily because any operator of the form $\mathcal{L}(dof_{i},dof_{j})$ tends to have strided memory access. ~\figref{fig:vector} shows the memory view for the global vector. 
Any loop over $(dof_{i})$ in vector assembly will write to memory in strided manner. For example, ($i = 0$) will write to 0,2 and  ($i = 1$)  to 1,3. This is not ideal for optimal performance. To circumvent this issue, we perform a sequence of the following operations:
\begin{enumerate}[label=\roman*]
    \item $zip$: We perform the $zip$ operations for the DOFs, where similar DOFs are arranged contiguously in memory. This is done by making just a single pass over the vector.
    \item Perform local vector assembly.
    \item $unzip$: Finally, we perform the $unzip$ operations to revert to the original global format.
\end{enumerate}
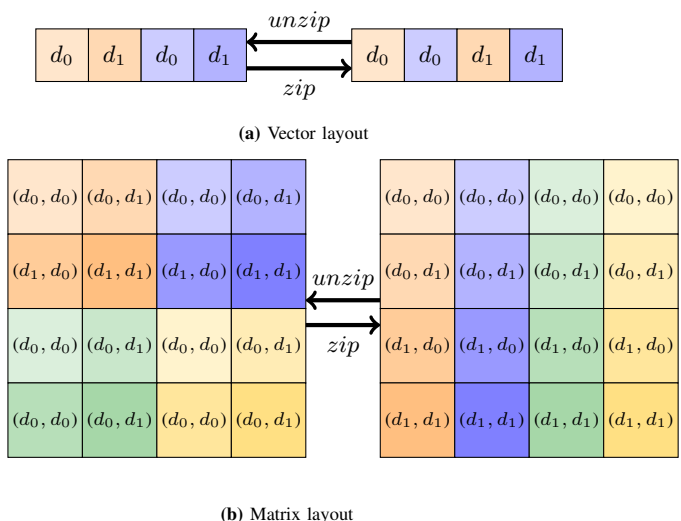
\begin{figure}
 \begin{subfigure}{0.9\linewidth}
\centering
    \begin{tikzpicture}[scale = 1.0]
    \def \dx{0.7}
    \def \dy{0.7}
    \foreach \x in {0}
        \draw[fill=orange!20] (\x,0) rectangle (\x+\dx,\dy)
        node[pos=.5] {\small{$d_{0}$}};
    \foreach \x in {\dx}
        \draw[fill=orange!30] (\x,0) rectangle (\x+\dx,\dy)
        node[pos=.5] {\small{$d_{1}$}};
    \foreach \x in {2*\dx}
        \draw[fill=blue!20] (\x,0) rectangle (\x+\dx,\dy)
        node[pos=.5] {\small{$d_{0}$}};
    \foreach \x in {3*\dx}
        \draw[fill=blue!30] (\x,0) rectangle  (\x+\dx,\dy)
        node[pos=.5] {\small{$d_{1}$}};
            \draw[->,ultra thick] (4*\dx,0.25*\dy) -- node[below] {\small{$zip$}} (6*\dx,0.25*\dy);
            \draw[<-,ultra thick] (4*\dx,0.75*\dy) -- node[above] {\small{$unzip$}} (6*\dx,0.75*\dy);
    \def \xspace{6*\dx}
          \foreach \x in {0}
        \draw[fill=orange!20] (\x + \xspace,0) rectangle  (\x+\dx+\xspace,\dy)
        node[pos=.5] {\small{$d_{0}$}};
    \foreach \x in {\dx}
        \draw[fill=blue!20] (\x+\xspace,0) rectangle  (\x+\dx+\xspace,\dy)
        node[pos=.5] {\small{$d_{0}$}};
    \foreach \x in {2*\dx}
        \draw[fill=orange!30] (\x+\xspace,0) rectangle  (\x+\dx+\xspace,\dy)
        node[pos=.5] {\small{$d_{1}$}};
     \foreach \x in {3*\dx}
        \draw[fill=blue!30] (\x+\xspace,0) rectangle  (\x+\dx+\xspace,\dy)
        node[pos=.5] {\small{$d_{1}$}};
    \end{tikzpicture}
 \caption{Vector layout}
 \label{fig:vector}
 \vspace{5 mm}
 \end{subfigure}
 \begin{subfigure}{0.85\linewidth}
\centering
    \begin{tikzpicture}[scale = 1.1]
\def \yspace{-0.5}
\def \dx{0.9}
\def \dy{0.9}
          \foreach \x in {0}
        \draw[fill=orange!20] 
        (\x,\yspace) rectangle (\x+\dx,\dy+\yspace)
        node[pos=.5] {\scriptsize{($d_{0},d_{0}$)}};
        \foreach \x in {\dx}
        \draw[fill=orange!30] 
        (\x,\yspace) rectangle (\x+\dx,\dy+\yspace)
        node[pos=.5] {\scriptsize{($d_{0},d_{1}$)}};
    \foreach \x in {2*\dx}
        \draw[fill=blue!20] 
        (\x,\yspace) rectangle (\x+\dx,\dy+\yspace)
        node[pos=.5] {\scriptsize{($d_{0},d_{0}$)}};
    \foreach \x in {3*\dx}
        \draw[fill=blue!30] 
        (\x,\yspace) rectangle (\x+\dx,\dy+\yspace)
        node[pos=.5] {\scriptsize{($d_{0},d_{1}$)}};
        \def \yspace{-1.4}
          \foreach \x in {0}
        \draw[fill=orange!40] 
        (\x,\yspace) rectangle (\x+\dx,\dy+\yspace)
        node[pos=.5] {\scriptsize{($d_{1},d_{0}$)}};
        \foreach \x in {\dx}
        \draw[fill=orange!50] 
        (\x,\yspace) rectangle (\x+\dx,\dy+\yspace)
        node[pos=.5] {\scriptsize{($d_{1},d_{1}$)}};
    \foreach \x in {2*\dx}
        \draw[fill=blue!40] 
        (\x,\yspace) rectangle (\x+\dx,\dy+\yspace)
        node[pos=.5] {\scriptsize{($d_{1},d_{0}$)}};
    \foreach \x in {3*\dx}
        \draw[fill=blue!50] 
        (\x,\yspace) rectangle (\x+\dx,\dy+\yspace)
        node[pos=.5] {\scriptsize{($d_{1},d_{1}$)}};
        
        \def \yspace{-2.3}
          \foreach \x in {0}
        \draw[fill=cpu1!20] 
        (\x,\yspace) rectangle (\x+\dx,\dy+\yspace)
        node[pos=.5] {\scriptsize{($d_{0},d_{0}$)}};
        \foreach \x in {\dx}
        \draw[fill=cpu1!30] 
        (\x,\yspace) rectangle (\x+\dx,\dy+\yspace)
        node[pos=.5] {\scriptsize{($d_{0},d_{1}$)}};
    \foreach \x in {2*\dx}
        \draw[fill=cpu2!20] 
        (\x,\yspace) rectangle (\x+\dx,\dy+\yspace)
        node[pos=.5] {\scriptsize{($d_{0},d_{0}$)}};
    \foreach \x in {3*\dx}
        \draw[fill=cpu2!30] 
        (\x,\yspace) rectangle (\x+\dx,\dy+\yspace)
        node[pos=.5] {\scriptsize{($d_{0},d_{1}$)}};
        
        \def \yspace{-3.2}
          \foreach \x in {0}
        \draw[fill=cpu1!40] 
        (\x,\yspace) rectangle (\x+\dx,\dy+\yspace)
        node[pos=.5] {\scriptsize{($d_{0},d_{0}$)}};
        \foreach \x in {\dx}
        \draw[fill=cpu1!50] 
        (\x,\yspace) rectangle (\x+\dx,\dy+\yspace)
        node[pos=.5] {\scriptsize{($d_{0},d_{1}$)}};
    \foreach \x in {2*\dx}
        \draw[fill=cpu2!40] 
        (\x,\yspace) rectangle (\x+\dx,\dy+\yspace)
        node[pos=.5] {\scriptsize{($d_{0},d_{0}$)}};
    \foreach \x in {3*\dx}
        \draw[fill=cpu2!50] 
        (\x,\yspace) rectangle (\x+\dx,\dy+\yspace)
        node[pos=.5] {\scriptsize{($d_{0},d_{1}$)}};
               \draw[->,ultra thick] (4*\dx,-1.6) -- node[below] {\small{$zip$}} (5*\dx,-1.6);
            \draw[<-,ultra thick] (4*\dx,-1.3) -- node[above] {\small{$unzip$}} (5*\dx,-1.3);
    \def \xspace{5*\dx}
    \def \yspace{-0.5}
      \foreach \x in {0}
        \draw[fill=orange!20] 
        (\x + \xspace,\yspace) rectangle (\x+\dx + \xspace,\dy+\yspace)
        node[pos=.5] {\scriptsize{($d_{0},d_{0}$)}};
        \foreach \x in {\dx}
        \draw[fill=blue!20]
         (\x + \xspace,\yspace) rectangle (\x+\xspace+\dx,\dy+\yspace)
        node[pos=.5] {\scriptsize{($d_{0},d_{0}$)}};
    \foreach \x in {2*\dx}
        \draw[fill=cpu1!20] 
         (\x+\xspace,\yspace) rectangle (\x+\dx+\xspace,\dy+\yspace)
        node[pos=.5] {\scriptsize{($d_{0},d_{0}$)}};
    \foreach \x in {3*\dx}
        \draw[fill=cpu2!20]  (\x+\xspace,\yspace) rectangle (\x+\dx+\xspace,\dy+\yspace)
        node[pos=.5] {\scriptsize{($d_{0},d_{0}$)}};
    \def \yspace{-1.4}
      \foreach \x in {0}
        \draw[fill=orange!30]   (\x + \xspace,\yspace) rectangle (\x+\dx+\xspace,\dy+\yspace)
        node[pos=.5] {\scriptsize{($d_{0},d_{1}$)}};
        \foreach \x in {\dx}
        \draw[fill=blue!30]  (\x+\xspace,\yspace) rectangle (\x+\dx+\xspace,\dy+\yspace)
        node[pos=.5] {\scriptsize{($d_{0},d_{1}$)}};
    \foreach \x in {2*\dx}
        \draw[fill=cpu1!30]  (\x+\xspace,\yspace) rectangle (\x+\dx+\xspace,\dy+\yspace)
        node[pos=.5] {\scriptsize{($d_{0},d_{1}$)}};
    \foreach \x in {3*\dx}
        \draw[fill=cpu2!30]  (\x+\xspace,\yspace) rectangle (\x+\dx+\xspace,\dy+\yspace)
        node[pos=.5] {\scriptsize{($d_{0},d_{1}$)}};
    \def \yspace{-2.3}
      \foreach \x in {0}
        \draw[fill=orange!40]  (\x+\xspace,\yspace) rectangle (\x+\dx+\xspace,\dy+\yspace)
        node[pos=.5] {\scriptsize{($d_{1},d_{0}$)}};
        \foreach \x in {\dx}
        \draw[fill=blue!40]  (\x+\xspace,\yspace) rectangle (\x+\dx+\xspace,\dy+\yspace)
        node[pos=.5] {\scriptsize{($d_{1},d_{0}$)}};
    \foreach \x in {2*\dx}
        \draw[fill=cpu1!40]  (\x+\xspace,\yspace) rectangle (\x+\dx+\xspace,\dy+\yspace)
        node[pos=.5] {\scriptsize{($d_{1},d_{0}$)}};
    \foreach \x in {3*\dx}
        \draw[fill=cpu2!40]  (\x+\xspace,\yspace) rectangle (\x+\dx+\xspace,\dy+\yspace)
        node[pos=.5] {\scriptsize{($d_{1},d_{0}$)}};
        
        \def \yspace{-3.2}
        \foreach \x in {0}
        \draw[fill=orange!50]  (\x+\xspace,\yspace) rectangle (\x+\dx+\xspace,\dy+\yspace)
        node[pos=.5] {\scriptsize{($d_{1},d_{1}$)}};
        \foreach \x in {\dx}
        \draw[fill=blue!50]  (\x+\xspace,\yspace) rectangle (\x+\dx+\xspace,\dy+\yspace)
        node[pos=.5] {\scriptsize{($d_{1},d_{1}$)}};
    \foreach \x in {2*\dx}
        \draw[fill=cpu1!50]  (\x+\xspace,\yspace) rectangle (\x+\dx+\xspace,\dy+\yspace)
        node[pos=.5] {\scriptsize{($d_{1},d_{1}$)}};
    \foreach \x in {3*\dx}
        \draw[fill=cpu2!50]  (\x+\xspace,\yspace) rectangle (\x+\dx+\xspace,\dy+\yspace)
        node[pos=.5] {\scriptsize{($d_{1},d_{1}$)}};

    \end{tikzpicture}
\caption{Matrix layout}
\label{fig:matrix}
 \end{subfigure}
 \vspace{1mm}
\caption{Vector memory layout  for a 2 DOF system in 1D: The left row depicts the memory layout where DOFs are written in strided memory. $zip$ operations zips all the DOFs together so that they are contiguous in memory whereas $unzip$ operation reverts back into the original memory layout} 
\end{figure}

Similar to vector assembly, matrix assembly faces the same issue. For example, in the case of 1D with 2 DOF system 
(\figref{fig:matrix}), 
$\mathcal{L}(dof_0,dof_0)$ would write to $0,2,8,10$ 
\footnote{considering flattened row major addressing}
. This leads to strided access. We perform a similar $zip$ operation to pack all the DOFs together, perform local matrix assembly and unzip to assemble into the global matrix.
A good way to look the $zip$ operation is that, after the $zip$ operation, the matrix size is $1\times4$, where ($dof_0,dof_0$) writes to $(0\cdots3)$, ($dof_0,dof_1$) to $(4\cdots8)$, ($dof_1,dof_0$) to $(9\cdots12)$ and ($dof_1,dof_1$) to $(13\cdots16)$. 

Additionally, we extended the idea proposed by \citet{saurabh2021industrial} by exposing each matrix assembly kernel as a matrix--matrix multiplication and vector assembly kernel as matrix--vector multiplication. This way of expressing the FEM operator allows us to leverage the benefits of vendor--optimized \textsc{GEMM} and \textsc{GEMV} kernels. We note the $zip$ and $unzip$ operations are critical to express the operator as matrix--matrix or matrix--vector multiplication. This also ensures the portability of the code across different hardware platforms. It forms a middle ground between portability and explicit vectorization of the FEM operator. 
\added{\subsection{Software design overview}}
\begin{figure}[]
    \centering
    \includegraphics[width=\linewidth]{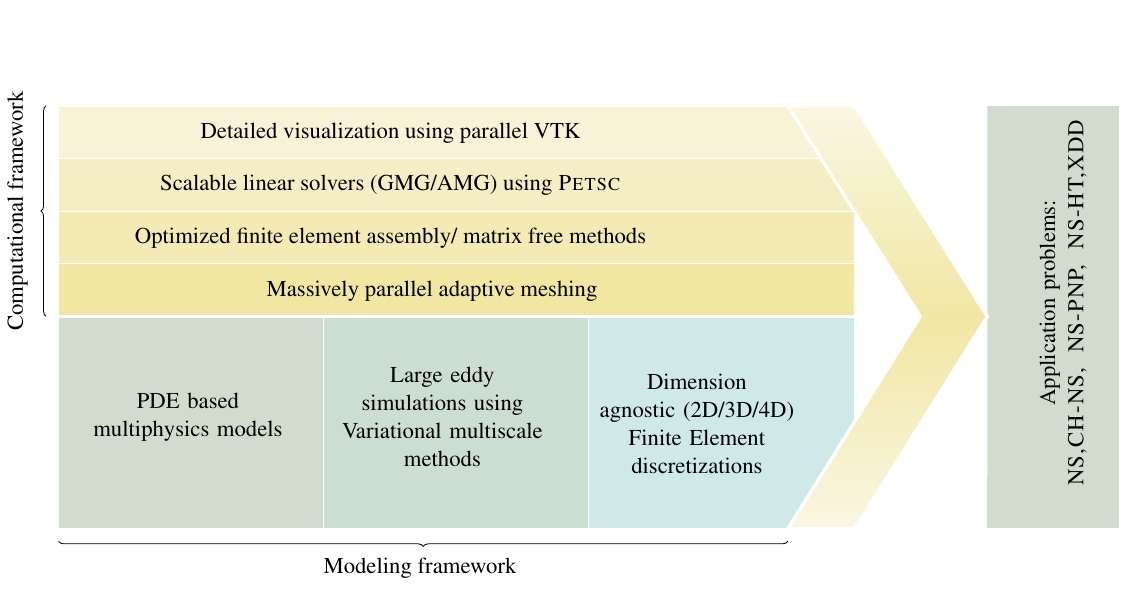}
    \caption{Brief software overview of the simulation framework}
    \label{fig:software}
\end{figure}

\added{\figref{fig:software} shows the brief overview of the software stack. The overall design consists of two major components -- modeling framework and computation framework. The modeling framework is comprised of an application-specific numerical discretization. The computational framework is comprised of a massively adaptive parallel octree-based framework for generating dimension-agnostic meshes. This is then interfaced with ~\petsc{} for scalable linear algebra. The dataset is output in a parallel VTK unstructured file format for detailed visualization using VTK compatible visualization software (\paraview~in our case). The layered design choice of the software stack makes the framework generalizable to various application problems and easier to maintain.}

\section{Results \& Discussion}

\subsection{Validation case - Swirling Flow}

\newcommand \figwidthCase{6.5}
\newcommand \figScale{0.11}
\begin{figure}[]
 \begin{center}
  \begin{tabular}{|c|c|c|c|}
  \hline
  & t = 0
  & t = 3.0
  & t = 5.5
  \\
  \hline
\rotatebox[origin=c]{90}{\footnotesize{$Cn = 0.0025$}
}
    & 
 \parbox[c]{\figwidthCase em}{
      {\includegraphics[width=0.11\textwidth,trim=6.8in 6.0in 2.5in 1.0in,clip
     ]
     {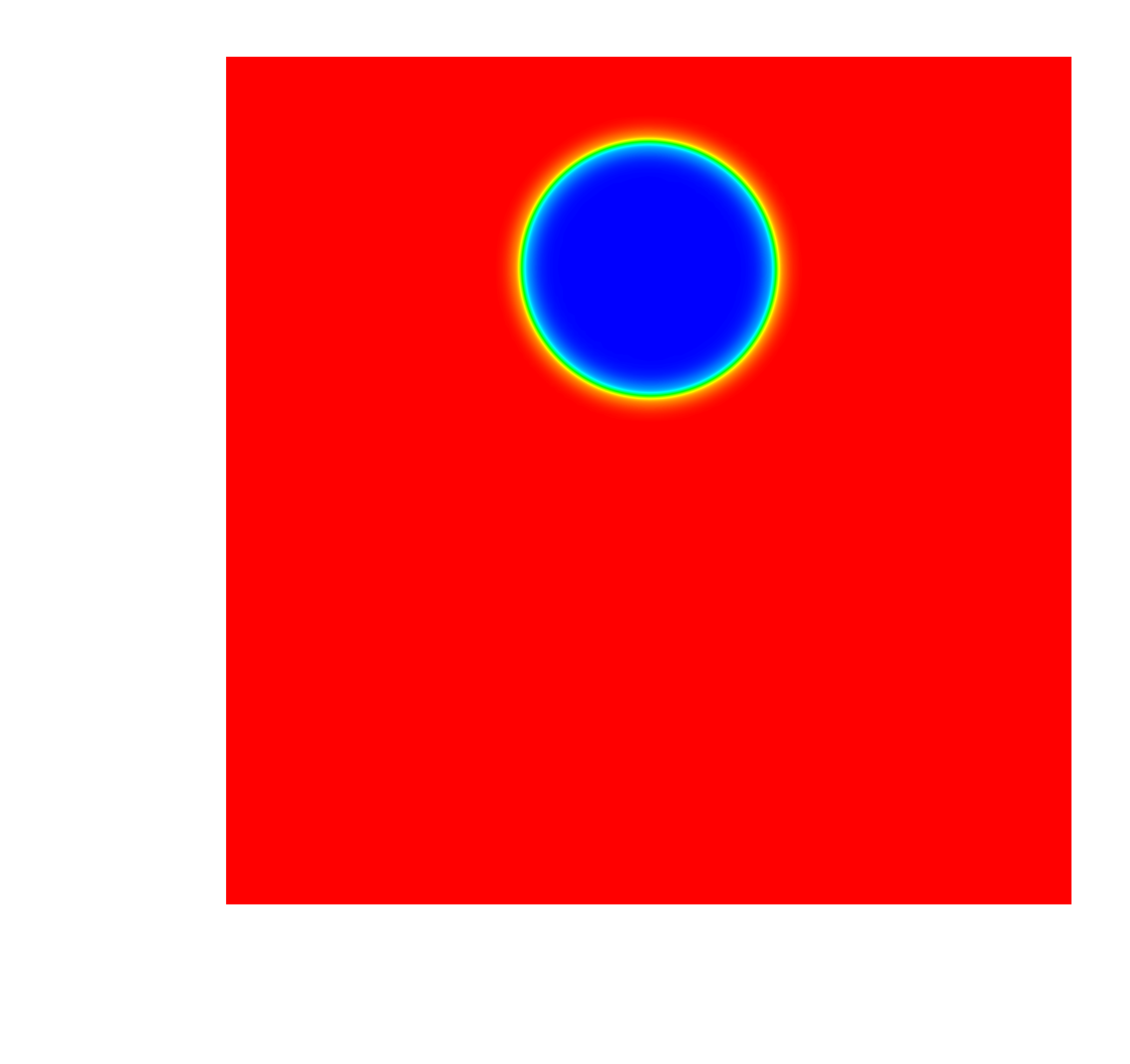}
 }}
      &
      \parbox[c]{\figwidthCase em}{
      \includegraphics[width=0.11\textwidth,trim=6.8in 6.0in 2.5in 1.0in,clip
      ]
      {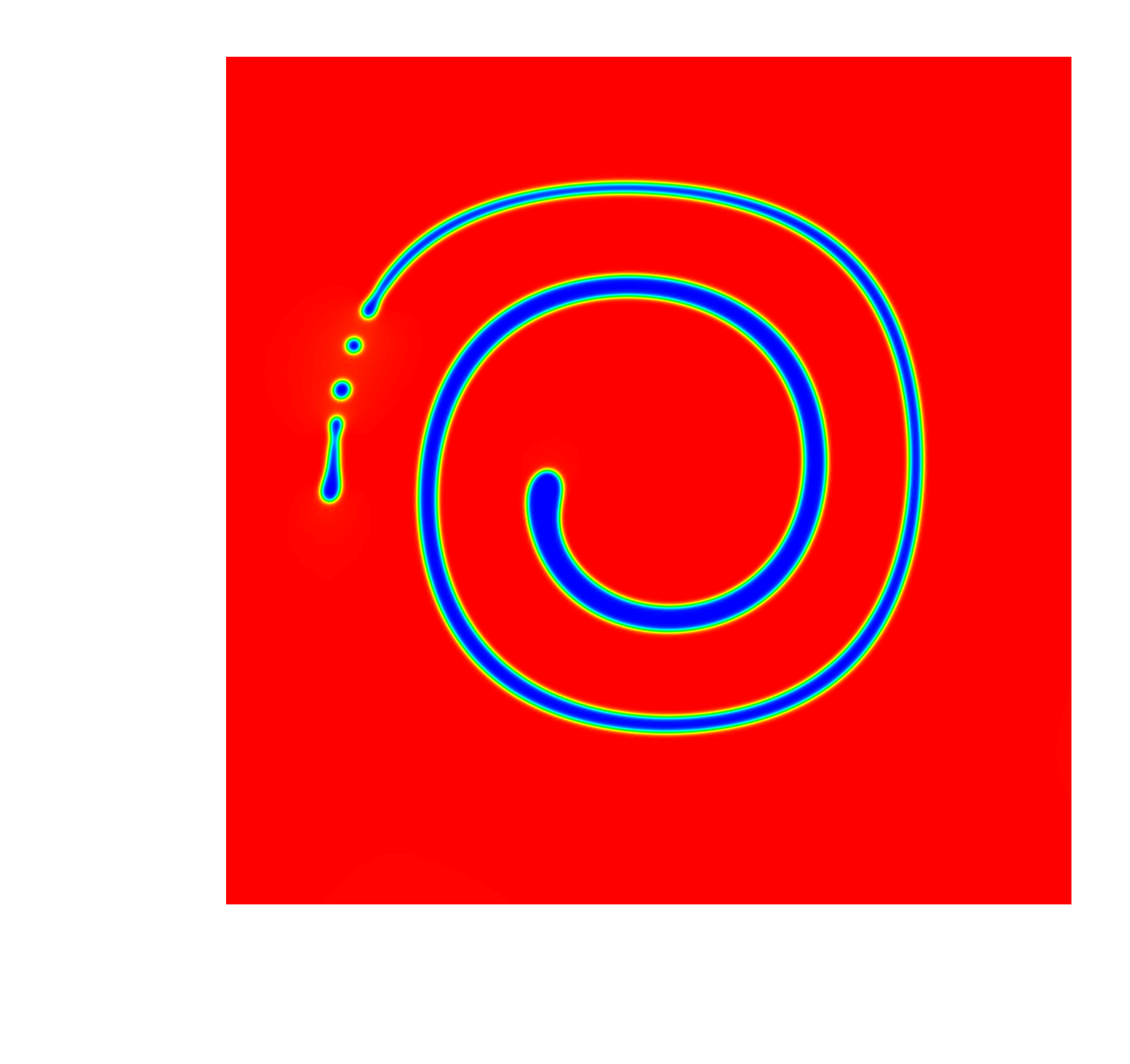}
  }
      &
      \parbox[c]{\figwidthCase em}{
      \includegraphics[width=0.11\textwidth,trim=6.8in 6.0in 2.5in 1.0in,clip
      ]
      {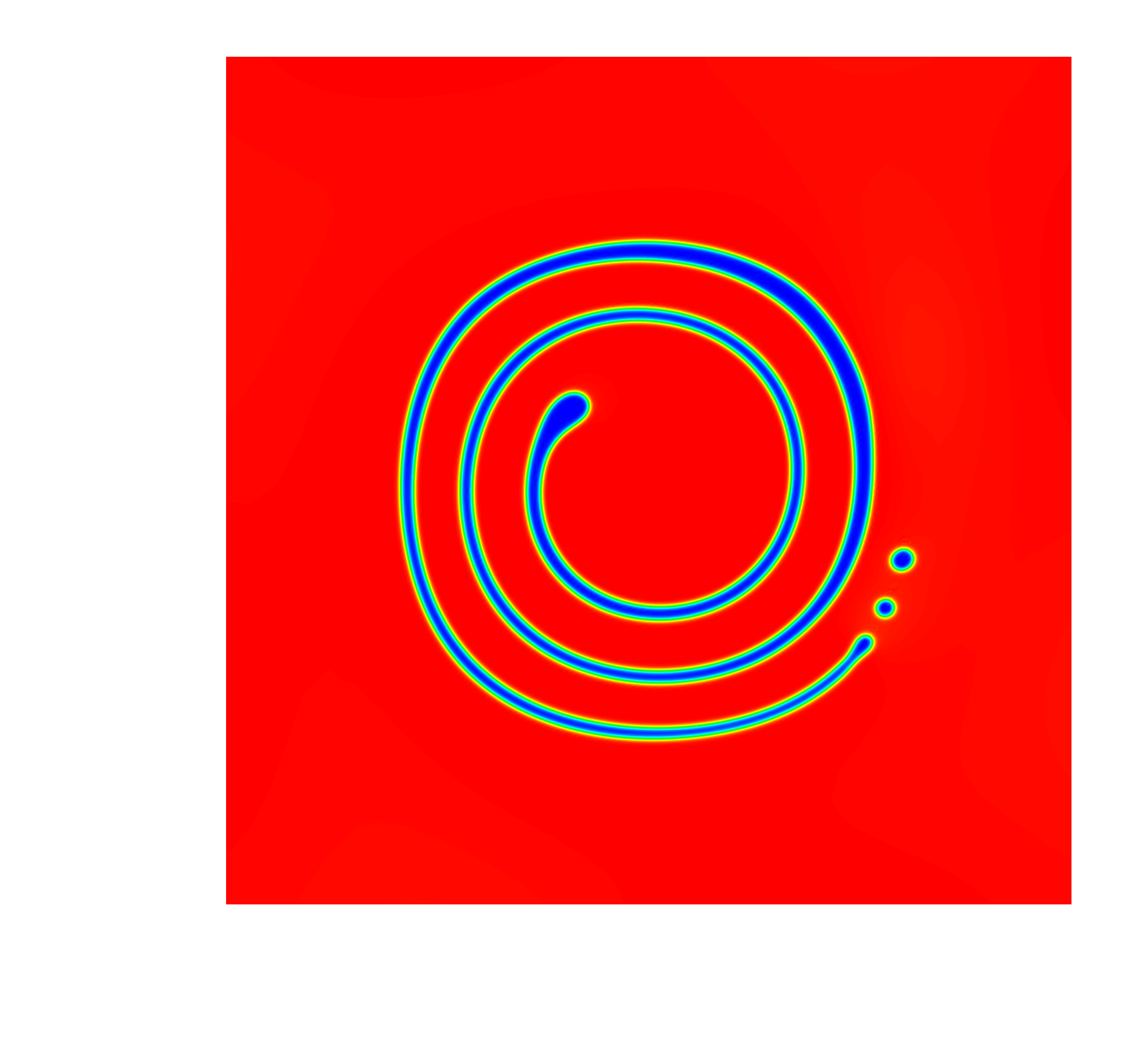}
  }
      \\
      \hline
     \rotatebox[origin=c]{90}{\footnotesize{$Cn = 0.001$}}
   & 
\parbox[c]{\figwidthCase em}{
     \includegraphics[width=0.11\textwidth,trim=6.8in 6.0in 2.5in 1.0in,clip
    ]{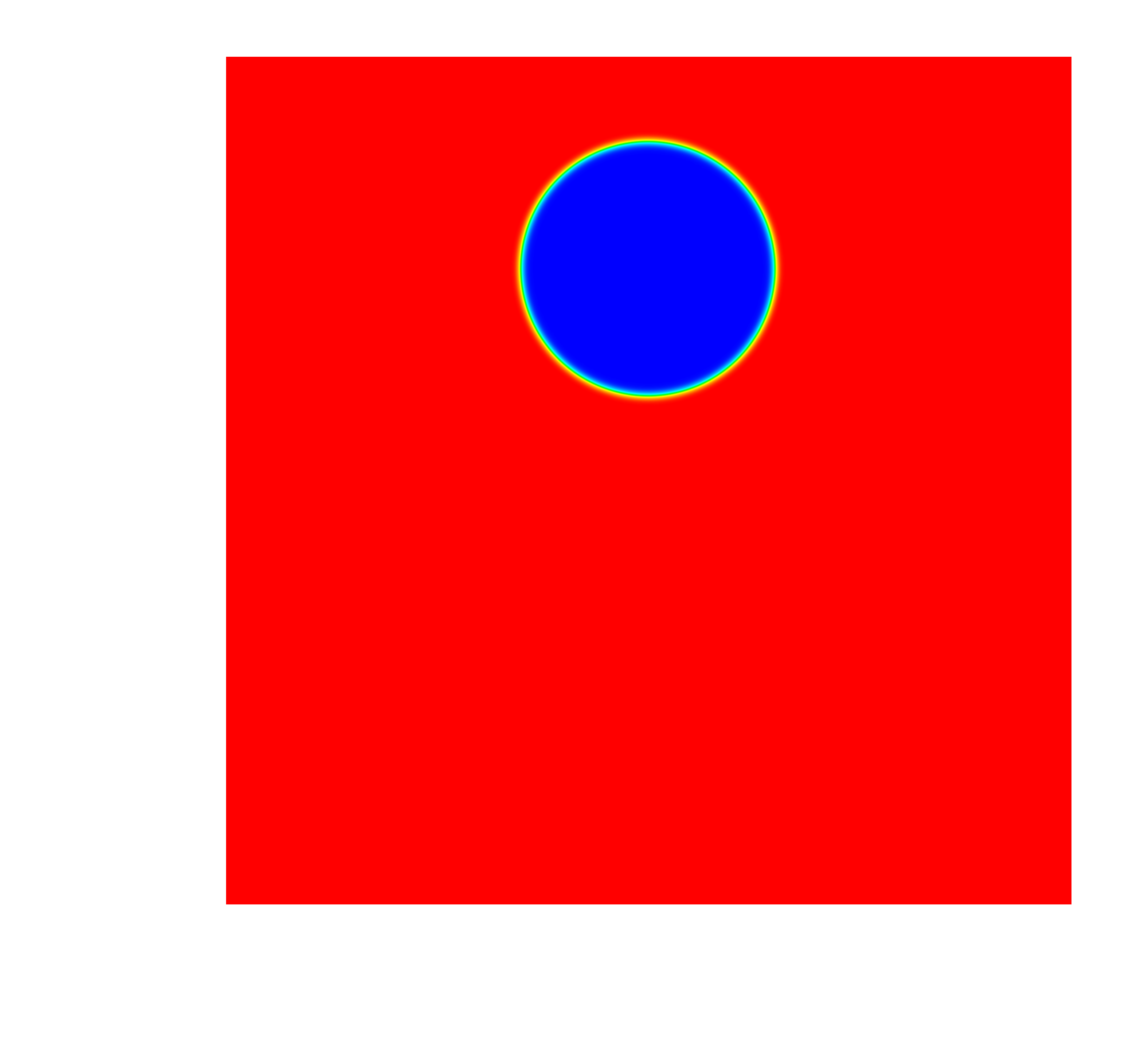}}
   & 
     \parbox[c]{\figwidthCase em}{
     \includegraphics[width=0.11\textwidth,trim=6.8in 6.0in 2.5in 1.0in,clip
     ]{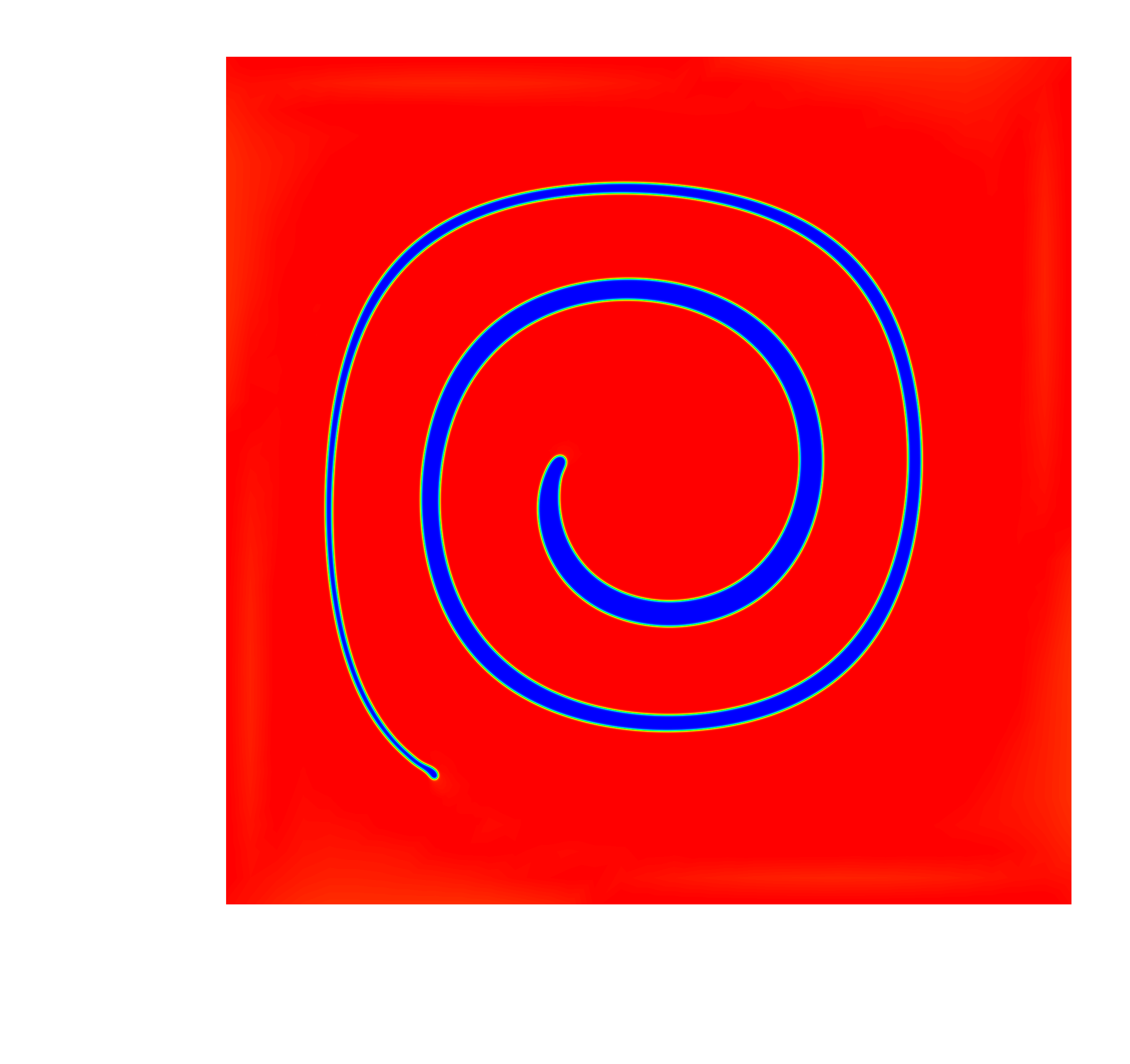}}
     &
     \parbox[c]{\figwidthCase em}{
     \includegraphics[width=0.11\textwidth,trim=6.8in 6.0in 2.5in 1.0in,clip
     ]{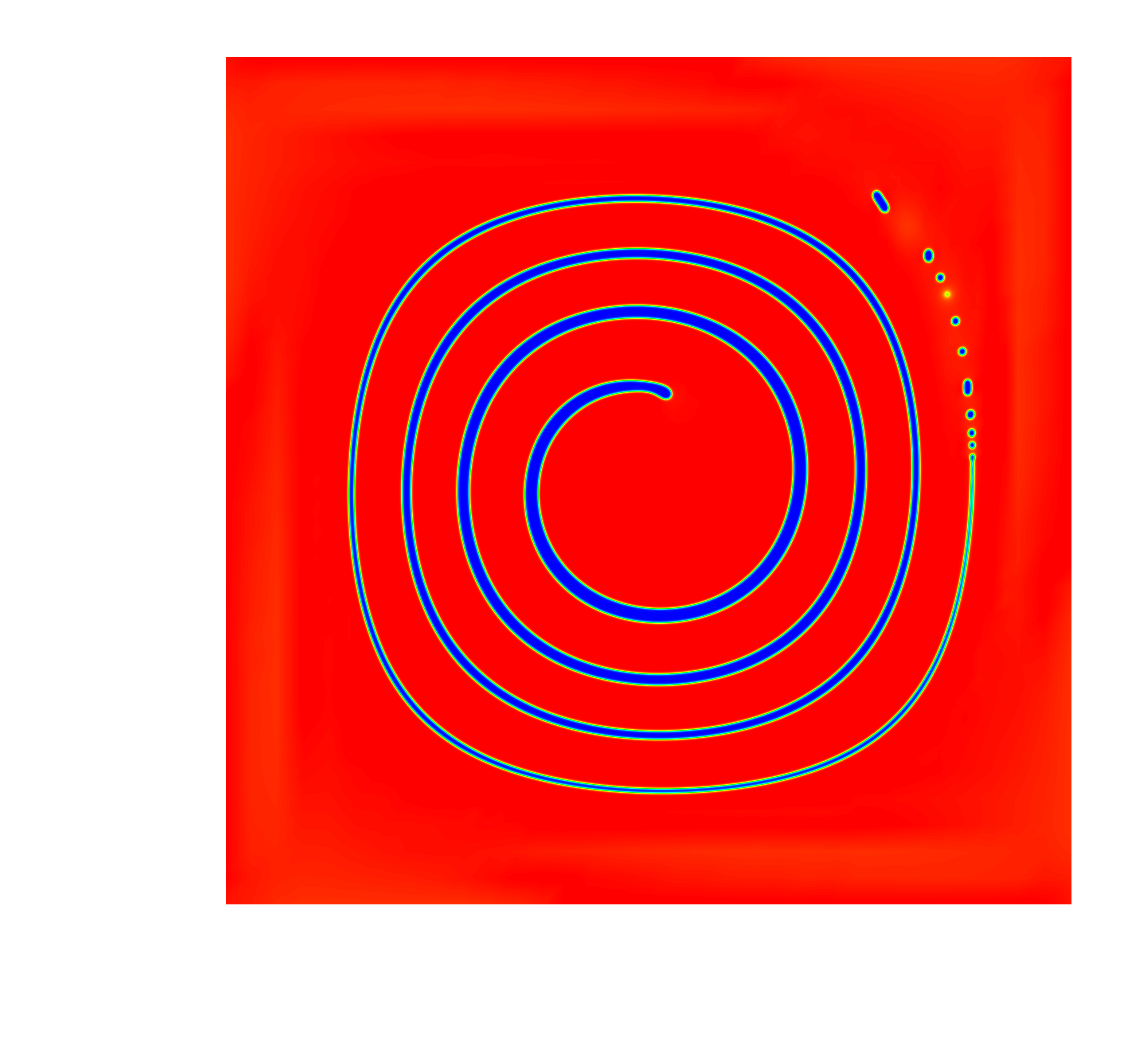}}
     \\
  
      \hline
      \rotatebox[origin=c]{90}{\footnotesize{local $Cn$}
      }
    & 
 \parbox[c]{\figwidthCase em}{
      \includegraphics[width=0.11\textwidth,trim=6.8in 6.0in 2.5in 1.0in,clip
     ]
     {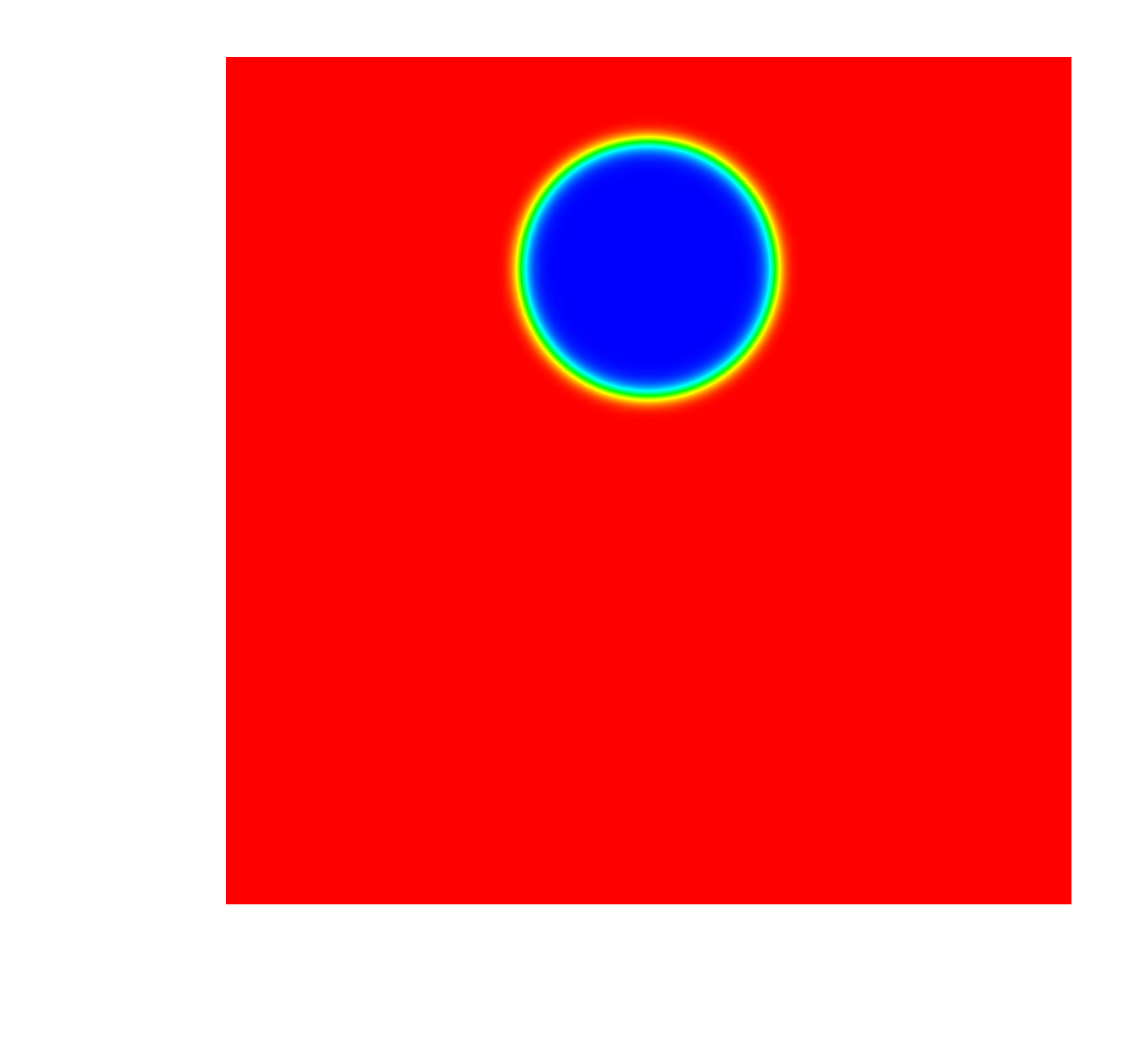}
 }
      &
      \parbox[c]{\figwidthCase em}{
      \includegraphics[width=0.11\textwidth,trim=6.8in 6.0in 2.5in 1.0in,clip
     ]
     {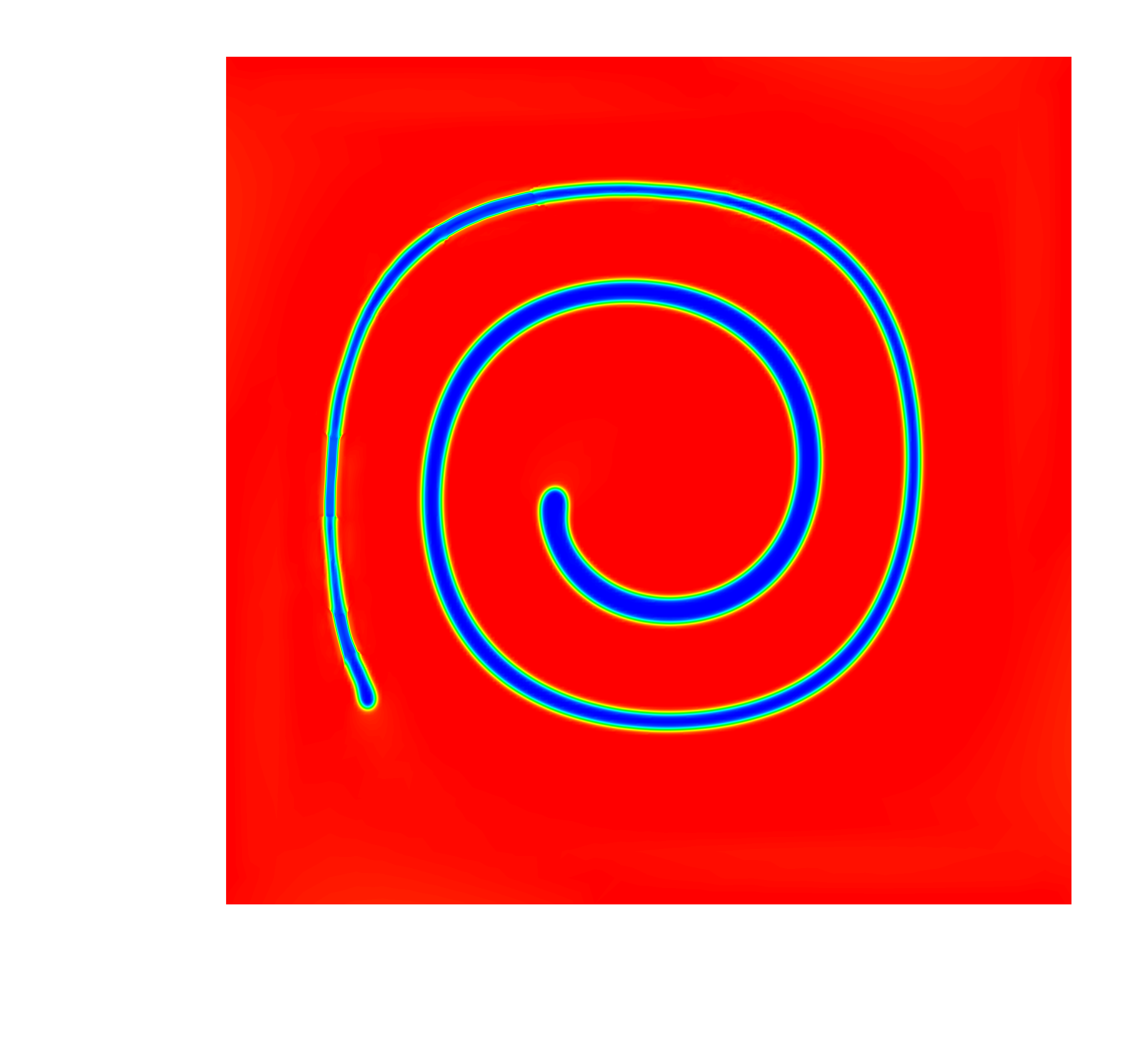}
 }
      &
      \parbox[c]{\figwidthCase em}{
      \includegraphics[width=0.11\textwidth,trim=6.8in 6.0in 2.5in 1.0in,clip
     ]
     {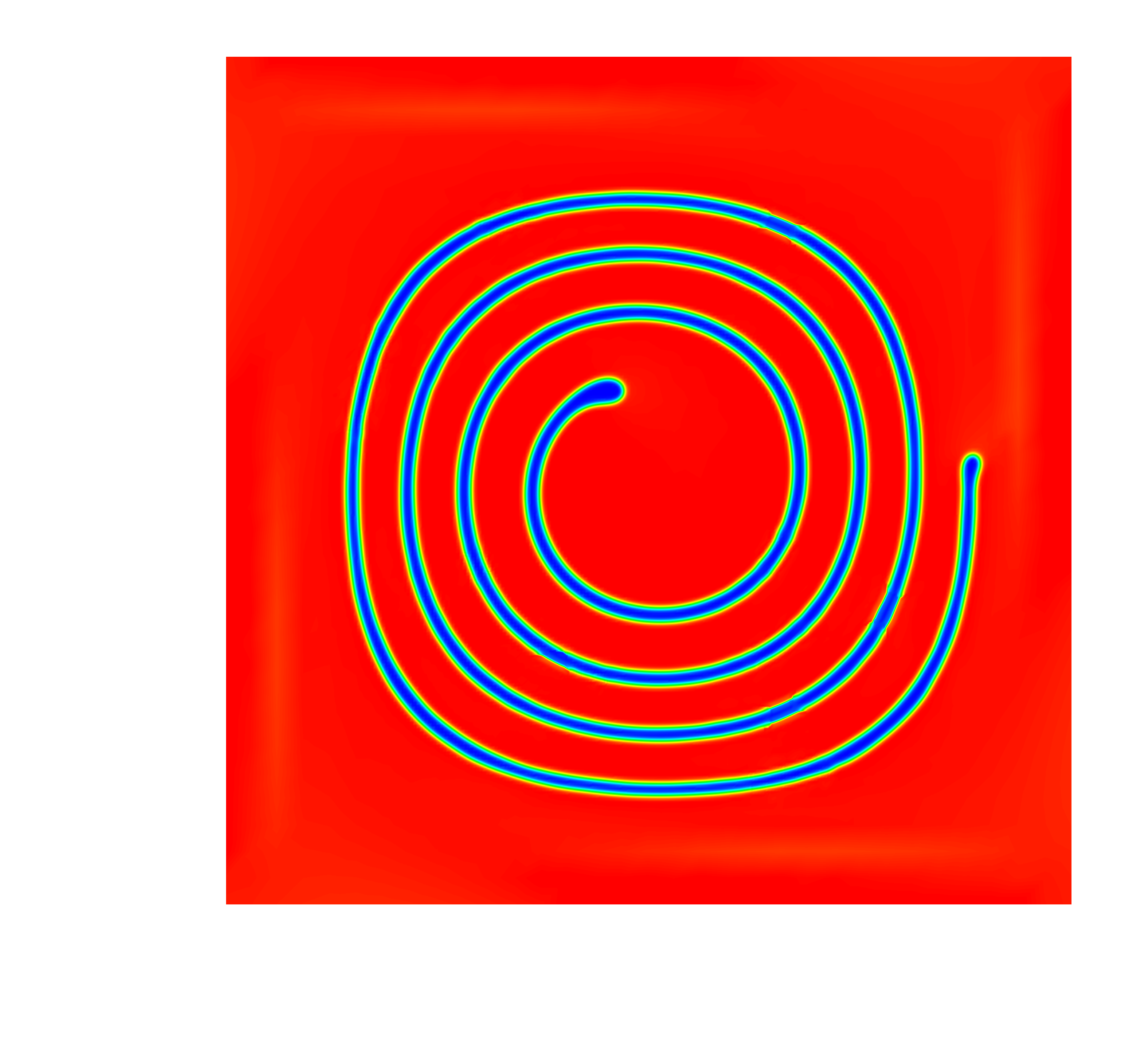}
 }
     \\
      \hline     
  \end{tabular}
  
  \end{center}
  \caption{\added{Time evolution for the swirling flow case. Top row plots results for (coarse) constant $Cn = 0.0025$ (interface refinement of level 9). Middle row plots results for (fine) constant $Cn = 0.001$ (interface refinement of level 12). 
  Bottom row shows the result of local $Cn$ where the interface is refined at level 9 with $Cn$ of 0.0025 and detected regions at level 12 with $Cn$ of 0.001.}}
  \label{fig:swirling_flow_time}
\end{figure}

\added{We illustrate the utility of the approach via a canonical example of a drop in swirling flow from~\citet{guo2022second}. A bubble of radius $0.15$ is located initially at (0.5, 75) in a non-dimensional domain of $1 \times 1$. The bubble is advected with a solenoidal velocity field having stream function $\psi = (1/\pi)  \sin^2(\pi x)\sin^2(\pi y)$. The velocity field drives the shape change of the drop into a thin spiraling filament. Capturing the dynamics of rapidly thinning filament represents a challenging test case.}


\added{  
\figref{fig:swirling_flow_time} compares the effect of local $Cn$ technique on capturing breakup dynamics. We compare three cases 1) (coarse) constant $Cn = 0.0025$, 2) (fine) constant $Cn = 0.001$, and 3) local $Cn$ where we start with a $Cn=0.0025$ and decrease it to $Cn = 0.001$ dynamically.     
For coarse $Cn = 0.025$ (top row), we observe numerical breakup due to insufficient resolution. The second row shows that this numerical breakup is minimized by decreasing $Cn = 0.001$ everywhere but at a much higher computational expense (level 12 instead of level 9 for $Cn = 0.0025$).  The third row shows results using the proposed local $Cn$ technique, where we perform a targeted decrease of $Cn$ to prevent numerical breakup. The bottom row produces very similar results to the middle row at a fraction of the computing cost (4 node hours on~\Stampede~compared to 44 node hours for $Cn = 0.001$). 
%
%
To the best of our knowledge, this is the first time anyone in the literature has reported a result across such a long time horizon (t = 5.5 non-dimensional time units) without any artificial breakup. We defer a detailed physics discussion of validation cases to our companion papers~\citep{Khanwale2023projection, dummy}.
}

\subsection{\added{Software optimization results}}

\added{We report the outcome of optimization carried out to boost the performance of the framework. We select a standard benchmark case of 3D bubble rise \cite{khanwale2020simulating} for reporting stage-by-stage improvement which is shown in ~\tabref{tab:optimization}. The interface is resolved at level 11, with bulk refinement at level 6. We evaluate across 11 timesteps. The simulation is carried out on 4 SKX nodes of TACC \Stampede{} (192 processes). The overall optimization is categorized into two stages:}
\begin{itemize}
    \item \added{Stage 1: The first stage of optimization involved incorporating multi-level remeshing along with the optimization for communication reported in \secref{sec:remesh-improvements}.  We observe an almost 2$\times$ improvement in the remeshing time as a result of this optimization. Additionally, \petsc~matrix data structure was converted from \texttt{MATMPIAIJ} to blocked format \texttt{MATBMPIAIJ}. This caused a significant reduction in the matrix assembly time for multi-dof equations, almost by a factor of 2$\times$ for CH solve (2 dof) and 2.2 $\times$ for velocity prediction (3 dof). Since pressure Poisson is a single dof equation, no significant improvement was observed. Additionally, we split the velocity update from a 3 dof solve to three single dof solves which reused the assembled matrix  (\secref{sec:LocalCahn}), but required three separate vector assemblies. The increase in the vector assembly cost is offset by the reduction in the matrix assembly cost resulting $\sim1.9\times$ improvement in VU solve.}

    \item \added{Stage 2: After stage 1 optimization, we observe that a significant amount of time is spent in the matrix and vector assembly (76\% for CH solve, 86\% for VP solve, 69.56\% for PP solve, and 87\% for VU solve). Therefore, we focused on optimizing the assembly routines. This is done by first zipping the consecutive dofs together, followed by exposing each operator in the FEM weak form as a matrix-vector (for vector assembly) or matrix-matrix-matrix product (for matrix assembly) rather than writing explicit loops over Gauss points. The matrix-vector and matrix-matrix products are performed by using the vendor-optimized (Intel MKL in present work) DGEMV and DGEMM kernel, respectively. As the final step, unzip operation is performed to cast the data back into the \petsc{} block format. Overall, we see a significant improvement in each of the solver stage as a result of this optimization, with CH showing a speedup of $1.4\times$, VP showing a speedup of $1.6\times$, PP showing a speedup of $1.58\times$ and VU a speedup of $2.08\times$ with respect to stage 1 optimization. The difference in the speedup observed for different stages can be attributed to the nature of FEM operator under consideration. We note that all the optimization performed here are hardware agnostic, but relies on the vendor optimized libraries for speedup, thereby ensuring the portability of the code across different platforms. }
\end{itemize}

\added{
{\renewcommand{\arraystretch}{1.1} 
\begin{table*}[h!]
\centering
{%
{
\begin{tabular}{|c|ccc|cccc|cccc|}
\hline
\multirow{2}{*}{} & \multicolumn{3}{c|}{\textbf{Baseline}}                                                       & \multicolumn{4}{c|}{\textbf{Stage 1}}                                                                                               & \multicolumn{4}{c|}{\textbf{Stage 2}}                                                                                               \\ \cline{2-12} 
                  & \multicolumn{1}{c|}{\textbf{Matrix}} & \multicolumn{1}{c|}{\textbf{Vector}} & \textbf{Total} & \multicolumn{1}{c|}{\textbf{Matrix}} & \multicolumn{1}{c|}{\textbf{Vector}} & \multicolumn{1}{c|}{\textbf{Total}} & \textbf{Improvement}   & \multicolumn{1}{c|}{\textbf{Matrix}} & \multicolumn{1}{c|}{\textbf{Vector}} & \multicolumn{1}{c|}{\textbf{Total}} & \textbf{Improvement}   \\ \hline
\textbf{CH}       & \multicolumn{1}{c|}{41.42}           & \multicolumn{1}{c|}{24.46}           & 91.68          & \multicolumn{1}{c|}{20.35}           & \multicolumn{1}{c|}{29.75}           & \multicolumn{1}{c|}{66.27}          & \textit{\textbf{1.38}} & \multicolumn{1}{c|}{15.20}           & \multicolumn{1}{c|}{14.36}           & \multicolumn{1}{c|}{45.41}          & \textit{\textbf{2.01}} \\ \hline
\textbf{VP}       & \multicolumn{1}{c|}{107.04}          & \multicolumn{1}{c|}{25.10}           & 154.52         & \multicolumn{1}{c|}{50.38}           & \multicolumn{1}{c|}{33.07}           & \multicolumn{1}{c|}{97.02}          & \textit{\textbf{1.59}} & \multicolumn{1}{c|}{35.30}           & \multicolumn{1}{c|}{12.26}           & \multicolumn{1}{c|}{60.70}          & \textit{\textbf{2.54}} \\ \hline
\textbf{PP}       & \multicolumn{1}{c|}{15.05}           & \multicolumn{1}{c|}{22.82}           & 50.40          & \multicolumn{1}{c|}{14.57}           & \multicolumn{1}{c|}{26.37}           & \multicolumn{1}{c|}{58.85}          & \textit{\textbf{0.86}} & \multicolumn{1}{c|}{10.88}           & \multicolumn{1}{c|}{8.26}            & \multicolumn{1}{c|}{37.02}          & \textit{\textbf{1.36}} \\ \hline
\textbf{VU}       & \multicolumn{1}{c|}{77.346}          & \multicolumn{1}{c|}{23.70}           & 119.67         & \multicolumn{1}{c|}{5.60}            & \multicolumn{1}{c|}{50.53}           & \multicolumn{1}{c|}{64.13}          & \textit{\textbf{1.86}} & \multicolumn{1}{c|}{3.39}            & \multicolumn{1}{c|}{21.14}           & \multicolumn{1}{c|}{30.80}          & \textit{\textbf{3.88}} \\ \hline
\textbf{Remesh}   & \multicolumn{2}{c|}{-}                                                      & 67.86          & \multicolumn{2}{c|}{-}                                                      & \multicolumn{1}{c|}{30.25}          & \textit{\textbf{2.19}} & \multicolumn{2}{c|}{-}                                                      & \multicolumn{1}{c|}{28.89}          & \textit{\textbf{2.24}} \\ \hline
\end{tabular}
}
}
\caption{\added{Comparison for the total time (in s) as a result of  code optimization for different stages of the solve. The improvement is measured as the speedup in the total time relative to the baseline timings.}}
\label{tab:optimization}
\end{table*}
}
}
\subsection{Region Identification: ~\mvec{} scaling}
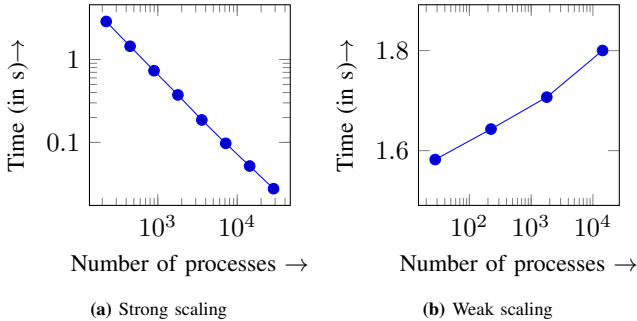
\begin{figure}
    \begin{subfigure}{0.48\linewidth}
    \begin{tikzpicture}
    \begin{loglogaxis}[width=\linewidth, height=\linewidth, scaled y ticks=true,
     xlabel={\small{Number of processes $\rightarrow$}},
    ylabel={\small{Time (in s)$\rightarrow$}},
		legend style={at={(0.5,-0.25)},anchor=north, nodes={scale=0.65, transform shape}}, 
		legend pos= south east, 
		legend columns=2,
		ytick={0.1,1.0},
		yticklabels={$0.1$,$1$},
		]
		\addplot table [x =npes,y expr=(\thisrow{ghostexchange}+\thisrow{matvec} ),col sep=space] {Data/StrongScalingMatVec.txt};
		\end{loglogaxis}
    \end{tikzpicture}
    \caption{Strong scaling}
    \label{fig:mavecStrongScaling}
    \end{subfigure}
    \begin{subfigure}{0.48\linewidth}
    \begin{tikzpicture}
    \begin{loglogaxis}[width=\linewidth, height=\linewidth, scaled y ticks=true,
    xlabel={\small{Number of processes $\rightarrow$}},
    ylabel={\small{Time (in s)$\rightarrow$}},
		legend style={at={(0.5,-0.25)},anchor=north, nodes={scale=0.65, transform shape}}, 
		legend pos= south east, 
		legend columns=2,
		ymin = 1.5,
        ymax = 1.9,
  		ytick={1.0,1.2,1.4,1.6,1.8},
		yticklabels={$1.0$,$1.2$,$1.4$,$1.6$,$1.8$},
		]
		\addplot table [x =npes,y expr=(\thisrow{ghostexchange}+\thisrow{matvec} ),col sep=space] {Data/WeakScalingMatVec.txt};
    \end{loglogaxis}
    \end{tikzpicture}
    \caption{Weak scaling}
    \label{fig:mavecWeakScaling}
    \end{subfigure}
    \vspace{2 mm}
    \caption{~\mvec{} scaling: Strong scaling and weak scaling for ~\mvec{} operation for linear basis function.}
    \label{fig:mvecscaling}
\end{figure}

The proposed algorithm for erosion and dilation scheme relies solely on the ~\mvec{} operations, which involve performing the local elemental traversal with associated ghost exchange. Therefore, we show the performance of ~\mvec{} operations of our framework for linear basis functions. ~\figref{fig:mvecscaling} shows the strong and weak scaling of ~\mvec{} for our framework. For strong scaling, we consider an adaptive mesh of around 13M elements and 13.7 M DOFs. \figref{fig:mavecStrongScaling} shows the ~\mvec{} execution time as a function of increasing number of cores. \mvec~ execution time decreased from 2.87 s on 224 processes to 0.027 s on 28K processes, resulting in 81\% parallel efficiency for 128 fold increase in processor count. For the weak scaling runs, we created grids with a fixed grain size of about 35K elements per core and timed \mvec~ execution time. The coarsest mesh consists of 981K elements on 28 processes  with 1.02M DOFs for linear and 8.01M DOFs.  \figref{fig:mavecWeakScaling} shows the  \mvec~ averaged over 100 iterations as a function of the number of cores. A constant execution time would imply ideal weak scaling efficiency. We observed a slowly growing weak-scaled execution time. Overall the time increased from about 1.58 s on 28 cores to 1.9 s on 14 K cores for linear elements, resulting in 82\% weak scaling efficiency.

\vspace{-0.2in}
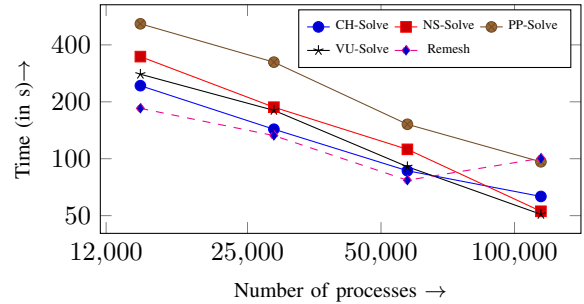
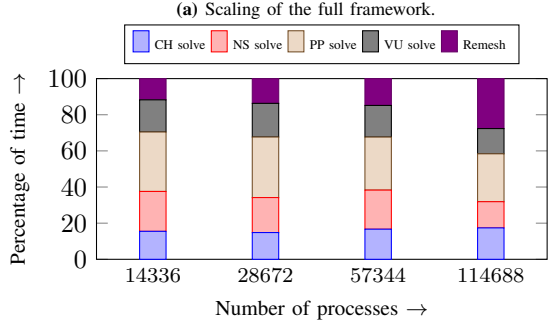
\begin{figure}
\centering
\begin{subfigure}{0.9\linewidth}
\begin{tikzpicture}       
    \begin{loglogaxis}[
        log basis x=10,
        log ticks with fixed point,
        width = \linewidth,
        height = 0.58\linewidth,
        xtick={6000,12000,25000,50000,100000},
        ytick={50,100,200,400,800},
xlabel = { \footnotesize Number of processes $\rightarrow$},
ylabel = {\footnotesize Time (in s)$\rightarrow$},
   legend style={legend columns=3},
        ]

	\addplot
 +[
 ]
	table[x expr={56*\thisrow{Nodes}))},y expr={\thisrow{Block1-ch} + \thisrow{Block2-ch}},col sep=space]{Data/scaling.txt};
	\addplot
 +[
 ]
	table[x expr={56*\thisrow{Nodes}))},y expr={\thisrow{Block1-ns} + \thisrow{Block2-ns}},col sep=space]{Data/scaling.txt};
	
	\addplot
 +[
 ]
	table[x expr={56*\thisrow{Nodes}))},y expr={\thisrow{Block1-pp} + \thisrow{Block2-pp}},col sep=space]{Data/scaling.txt};
		
    \addplot
    +[
    ]
	table[x expr={56*\thisrow{Nodes}))},y expr={\thisrow{Block1-vu} + \thisrow{Block2-vu}},col sep=space]{Data/scaling.txt};
	
	\addplot
 +[
 dashed,
 magenta,
 ]
	table[x expr={56*\thisrow{Nodes}))},y expr={\thisrow{Remesh}},col sep=space]{Data/scaling.txt};

	\legend{\tiny{CH-Solve}, \tiny{NS-Solve},\tiny{PP-Solve},\tiny{VU-Solve}, \tiny{Remesh}}
   \end{loglogaxis}
\end{tikzpicture}
\caption{Scaling of the full framework.}
\label{fig:scalingTime}

\end{subfigure}
\begin{subfigure}{0.9\linewidth}
\vspace{1em}
    \begin{tikzpicture}
    \begin{axis}[
        width=0.95\linewidth,
        height=0.50\linewidth,
        ybar stacked,
        ymin=0,
        ymax=100,
        xticklabel style={rotate=0,font=\footnotesize},
        xticklabels = {$14336$,$28672$,$57344$,$114688$}, %
        xtick=data, 
        enlarge x limits={abs=0.5},
        ylabel={\footnotesize Percentage of time $\rightarrow$},
        xlabel={\footnotesize Number of processes $\rightarrow$},
        legend style={at={(0.5,1.3)},anchor= north,legend columns=5}, 
        ]
        \addplot table[x expr=\coordindex,y expr={(\thisrow{Block1-ch} + \thisrow{Block2-ch})*100/(\thisrow{Block1-ch} + \thisrow{Block2-ch} + \thisrow{Block1-vu} + \thisrow{Block2-vu} + \thisrow{Block1-pp} + \thisrow{Block2-pp} + \thisrow{Block1-ns} + \thisrow{Block2-ns} + \thisrow{Remesh})},col sep=space] {Data/scaling.txt};
        
        \addplot table[x expr=\coordindex,y expr={(\thisrow{Block1-ns} + \thisrow{Block2-ns})*100/(\thisrow{Block1-ch} + \thisrow{Block2-ch} + \thisrow{Block1-vu} + \thisrow{Block2-vu} + \thisrow{Block1-pp} + \thisrow{Block2-pp} + \thisrow{Block1-ns} + \thisrow{Block2-ns} + \thisrow{Remesh})},col sep=space] {Data/scaling.txt};

        \addplot table[x expr=\coordindex,y expr={(\thisrow{Block1-pp} + \thisrow{Block2-pp})*100/(\thisrow{Block1-ch} + \thisrow{Block2-ch} + \thisrow{Block1-vu} + \thisrow{Block2-vu} + \thisrow{Block1-pp} + \thisrow{Block2-pp} + \thisrow{Block1-ns} + \thisrow{Block2-ns} + \thisrow{Remesh})},col sep=space] {Data/scaling.txt};

        \addplot table[x expr=\coordindex,y expr={(\thisrow{Block1-vu} + \thisrow{Block2-vu})*100/(\thisrow{Block1-ch} + \thisrow{Block2-ch} + \thisrow{Block1-vu} + \thisrow{Block2-vu} + \thisrow{Block1-pp} + \thisrow{Block2-pp} + \thisrow{Block1-ns} + \thisrow{Block2-ns} + \thisrow{Remesh})},col sep=space] {Data/scaling.txt};
        
    \addplot table[x expr=\coordindex,y expr={\thisrow{Remesh})*100/(\thisrow{Block1-ch} + \thisrow{Block2-ch} + \thisrow{Block1-vu} + \thisrow{Block2-vu} + \thisrow{Block1-pp} + \thisrow{Block2-pp} + \thisrow{Block1-ns} + \thisrow{Block2-ns} + \thisrow{Remesh})},col sep=space] {Data/scaling.txt};

        \legend{\tiny CH solve, \tiny NS solve, \tiny PP solve, \tiny VU solve, \tiny Remesh}
    \end{axis}
    \end{tikzpicture}
    \caption{\added{Percentage of time for individual steps.}}
    \label{fig:scalingRelCost}
\end{subfigure}   
\vspace{0.5em}
\caption{\textbf{Application Scaling}: Scaling behavior of the overall framework (\figref{fig:scalingTime}) and relative cost (\figref{fig:scalingRelCost}) of each individual component for 700M mesh.  }
\label{fig:applicationScaling}
\end{figure}
\subsection{Application performance scaling}

\begin{figure*}
	\includegraphics[width=\linewidth]{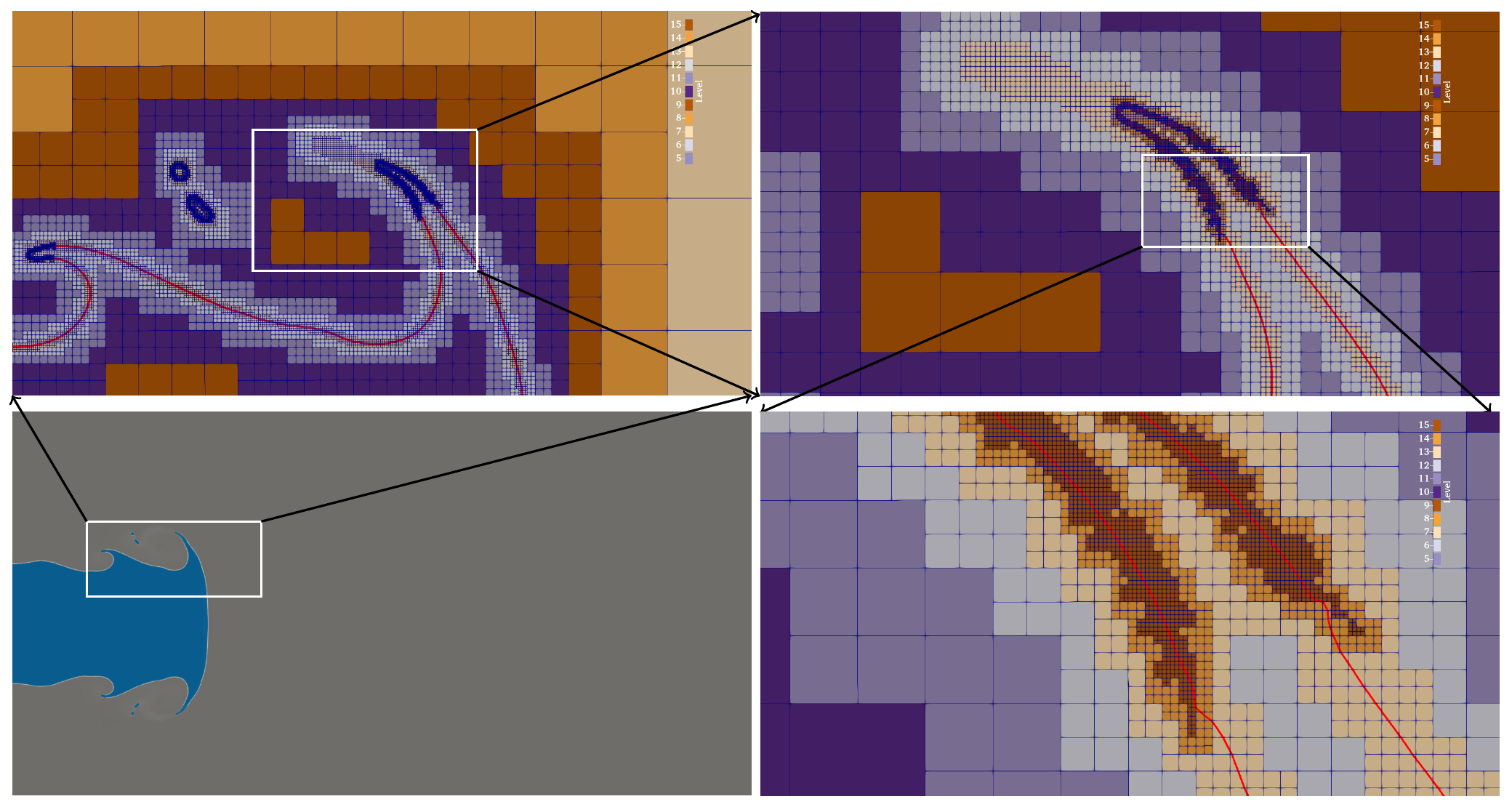}
	\caption{\textbf{Adaptive mesh refinement}: 2D slice with mesh overlayed. The interface is marked by dark red color. Note the tip of the filament (in the bottom right figure) and small bubble (top left) is much more resolved then the other regions of the interface. This region is identified by the erosion and dilation algorithm described in ~\secref{sec:FeatueDetection}. The octree level differs by 10 levels, with coarsest mesh at level 5 and the finest at level 15, resulting in a $10^9 \times$ difference in the elemental volume between the finest and the coarsest elements.}
	\label{fig:zoomview}
\end{figure*}

\begin{table}[h]
\centering
\begin{tabular}{|c|c|c|c|c|}
\hline
 & \textbf{CH solve} & \textbf{NS Solve} & \textbf{PP solve} & \textbf{VU solve} \\ \hline
\textbf{Solver} & bcgs & bcgs & ibcgs & cg \\ \hline
\textbf{PCType} & bjacobi & bjacobi & bjacobi & jacobi \\ \hline
\end{tabular}
\caption{\added{Solver and preconditioner for the solvers.}}
\label{tab:solver}
\end{table}
~\figref{fig:applicationScaling} shows the scaling performance of the full framework on TACC ~\Frontera. 
\footnote{All simulation was performed using 56 cores per node}. 
The overall mesh consists of around 700 M mesh elements. The simulation was run for 11 timesteps. We timed each of the solvers and remeshing steps separately. \added{\tabref{tab:solver} specifies the linear solver used along with the preconditioner used for the various stages. The absolute and relative tolerances were set to \num{e-8} for the linear solve and \num{e-10} for the non-linear solve, respectively.} Overall, we observe good scaling behavior for all solvers (\figref{fig:scalingTime}) with NS-solve resulting in $6.6 \times$ speedup for $8 \times$ increase in the processor from $\mathcal{O}(14K)$ processor to $\mathcal{O}(112K)$ processor. Similarly, PP--solve showed a $5.3 \times$ reduction, whereas VU--solve showed $5.5\times$ and CH--solve showed a $4 \times$ reduction in total solve time. The remeshing cost showed $2.5 \times$ reduction for a $4 \times$ increase in processes till $\mathcal{O}(57K)$ processes but then began to grow with further increase in processor. We suspect that this increased cost is due to communication (for this grain size) and may be alleviated for larger problems and by using hybrid parallelism. This remains an open problem we are currently pursuing. 

The variable density PP--solve step is the most time-consuming step until remeshing cost begins to dominate (\figref{fig:scalingRelCost}). Solving pressure Poisson efficiently, especially with variable coefficients, is still an active area of research. Scalable solvers, like Geometric multigrid (GMG), promise to yield a better solve time but rely on optimized algorithms for creating different mesh hierarchies and ~\mvec{} operation. 
This is left as future work.
\footnote{We have tested the framework with Algebraic multigrid via ~\petsc{} and Hypre interface. Our study showed that the Krylov space solver yields a better solve time at a large processor count due to the setup cost associated with AMG. Therefore, we used iterative solvers for the current study.}

\textbf{Weak scaling:} We do not produce any weak scaling results as it is hard to create an adaptive mesh with a fixed grain size per processor. Arbitrarily refining the elements lead to a much-increased iteration count during the KSP solve and is therefore not a correct metric for the scaling study for a real-world application.

\section{Application to primary jet atomization}
\label{sec:Atomization}
We finally demonstrate the ability of our algorithm to simulate the most resolved simulation of primary jet atomization. The finest resolution in our application problem consists of octree at level 15. This is equivalent to solving it on a 35 trillion grid point on a uniform mesh - a 64$\times$ larger than the current state-of-the-art~\citep{pairetti2020mesh}. \figref{fig:jet} shows the simulation result at \SI{6.3}{\micro\metre}. We can see that the simulation framework can capture tiny droplets by selectively identifying the key regions of interest and selectively increasing the resolution.

\begin{figure}
    \centering
\begin{tikzpicture}
\begin{axis}[
    width=0.9\linewidth,
    height=0.5\linewidth,
    ybar ,
    ymin=0,
    xticklabels = {$\leq10$,$11$,$12$,$13$,$14$,$15$}, %
    xtick=data, 
    enlarge x limits=0.05, 
    ylabel= \footnotesize{Element fraction},
    xlabel= \footnotesize{Level },
    ]
     \addplot table[x expr=\coordindex,y =ElementFrac] {Data/ElementDistributionReduced.txt}; 
\end{axis}
\end{tikzpicture}    
\caption{Variation of the element fraction with respect to the levels.}
\label{fig:elemFraction}
\end{figure}
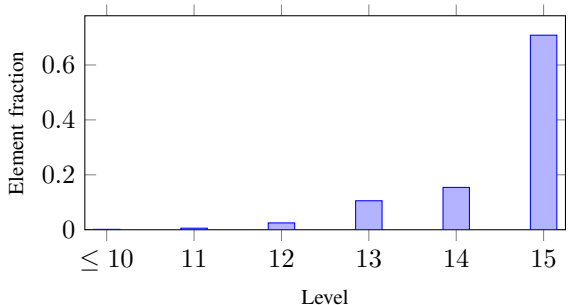
\figref{fig:zoomview} shows the progressive refinement of the mesh. We can see that the algorithm proposed in ~\secref{sec:FeatueDetection} can detect complicated structures like filaments and drops and selectively refine those regions. The overall interface is resolved at the level
13, with the key features resolved at level 15. ~\figref{fig:elemFraction} shows the fraction  of the element at different levels. We see that a significant portion of the elements are at levels 13 and 14 ($\approx 25 \%)$, with a maximum fraction at level 15. A rough estimate indicates that resolving the complete interface at level 15 would result in at least an $8 - 10 \times$ increase in the overall mesh count. This would result in  $20-25\times$ higher time to solve,  
\footnote{assuming $\mathcal{O}(NlogN)$ scaling of KSP solve.} 
making such simulations impractical in a reasonable time. We note that level 15 has the maximum element fraction but covers only $0.01\%$ of total volume. This illustrates the importance of adaptivity in resolving multiphase simulations. We defer a detailed physics discussion of this simulation to our companion physics paper~\citep{dummy}.
\section{Conclusion \& Future work}
In this work, we present algorithmic advances in the field of multiphase flow modeling and simulation. We propose a  scalable algorithm to identify the key regions of interest that need sufficiently high resolution for capturing the flow physics accurately. This approach is critical for efficient and accurate simulation of multiphase flows in a reasonable time. Further, we modify the octree refinement algorithm to refine to an arbitrary level in a single step. We also accelerate the matrix and vector assembly by carefully analyzing the data movement. We showcase the application of these algorithmic advances by simulating the canonical problem of primary jet atomization. These algorithmic advances enable us to push the current state--of--the--art by a factor of 64$\times$ in terms of the resolved physics. We believe the advances proposed in this work can serve as a reference for the next--generation of multiphase flow simulations.

In future, we plan to extend these algorithms, by integrating the framework with GPU accelerated linear algebra solvers like \textsc{Amat}~\cite{tran2022scalable}, \textsc{Magma}~\cite{tomov2009magma}. Further, we plan to improve the remeshing scaling and utilize GMG to improve the solve--time, specifically for the variable coefficient pressure Poisson problem. We also plan to integrate the setup with Domain Specific Language (DSL) like \textsc{Finch}~\cite{heisler2022domain}, \textsc{Fenics}~\cite{alnaes2015fenics} and extend FEM framework to Finite Volume and Discontinuous Galerkin schemes.

\vspace{-2 mm}
\section{Software availabilty}
The code called \texttt{Proteus} is open-source and available under the GPL V3.0 License at:  \href{https://bitbucket.org/baskargroup/proteus/src/master/}{https://bitbucket.org/baskargroup/proteus/src/master/}. 

\section*{Acknowledgement}
We acknowledge NSF LEAP-HI award number 2053760 and NSF CPS Frontiers award number 1954556, the AI Research Institutes program supported by NSF and USDA-NIFA under AI Institute: for Resilient Agriculture, Award No. 2021-67021-35329, OAC 1808652.  We are grateful to TACC for computing resources on Frontera through award number ASC21024.  We are also grateful to XSEDE for computing resources on Stampede2, Bridges2, and Expanse through award number CTS110007.
\bibliography{main}

\begin{thebibliography}{34}
\providecommand{\natexlab}[1]{#1}
\providecommand{\url}[1]{#1}
\csname url@samestyle\endcsname
\providecommand{\newblock}{\relax}
\providecommand{\bibinfo}[2]{#2}
\providecommand{\BIBentrySTDinterwordspacing}{\spaceskip=0pt\relax}
\providecommand{\BIBentryALTinterwordstretchfactor}{4}
\providecommand{\BIBentryALTinterwordspacing}{\spaceskip=\fontdimen2\font plus
\BIBentryALTinterwordstretchfactor\fontdimen3\font minus
  \fontdimen4\font\relax}
\providecommand{\BIBforeignlanguage}[2]{{%
\expandafter\ifx\csname l@#1\endcsname\relax
\typeout{** WARNING: IEEEtranN.bst: No hyphenation pattern has been}%
\typeout{** loaded for the language `#1'. Using the pattern for}%
\typeout{** the default language instead.}%
\else
\language=\csname l@#1\endcsname
\fi
#2}}
\providecommand{\BIBdecl}{\relax}
\BIBdecl

\bibitem[Joshi and Nandakumar(2015)]{joshi2015computational}
J.~Joshi and K.~Nandakumar, ``Computational modeling of multiphase reactors,''
  \emph{Annual review of chemical and biomolecular engineering}, vol.~6, pp.
  347--378, 2015.

\bibitem[Herrmann(2010{\natexlab{a}})]{Herrmann2010}
M.~Herrmann, ``Detailed numerical simulations of the primary atomization of a
  turbulent liquid jet in crossflow,'' \emph{Journal of Engineering for Gas
  Turbines and Power}, vol. 132, no.~6, p. 061506, 2010.

\bibitem[Fuster et~al.(2009)Fuster, Bagu{\'e}, Boeck, Le~Moyne, Leboissetier,
  Popinet, Ray, Scardovelli, and Zaleski]{fuster2009simulation}
D.~Fuster, A.~Bagu{\'e}, T.~Boeck, L.~Le~Moyne, A.~Leboissetier, S.~Popinet,
  P.~Ray, R.~Scardovelli, and S.~Zaleski, ``Simulation of primary atomization
  with an octree adaptive mesh refinement and vof method,'' \emph{International
  Journal of Multiphase Flow}, vol.~35, no.~6, pp. 550--565, 2009.

\bibitem[Tomar et~al.(2010)Tomar, Fuster, Zaleski, and
  Popinet]{tomar2010multiscale}
G.~Tomar, D.~Fuster, S.~Zaleski, and S.~Popinet, ``Multiscale simulations of
  primary atomization,'' \emph{Computers \& Fluids}, vol.~39, no.~10, pp.
  1864--1874, 2010.

\bibitem[Pairetti et~al.(2020)Pairetti, Dami{\'a}n, Nigro, Popinet, and
  Zaleski]{pairetti2020mesh}
C.~I. Pairetti, S.~M. Dami{\'a}n, N.~M. Nigro, S.~Popinet, and S.~Zaleski,
  ``Mesh resolution effects on primary atomization simulations,''
  \emph{Atomization and Sprays}, vol.~30, no.~12, 2020.

\bibitem[Herrmann(2010{\natexlab{b}})]{herrmann2010detailed}
M.~Herrmann, ``Detailed numerical simulations of the primary atomization of a
  turbulent liquid jet in crossflow,'' \emph{Journal of Engineering for Gas
  Turbines and Power}, vol. 132, no.~6, 2010.

\bibitem[Gorokhovski and Herrmann(2008)]{gorokhovski2008modeling}
M.~Gorokhovski and M.~Herrmann, ``Modeling primary atomization,'' \emph{Annu.
  Rev. Fluid Mech.}, vol.~40, pp. 343--366, 2008.

\bibitem[Lu and Tryggvason(2018)]{lu2018direct}
J.~Lu and G.~Tryggvason, ``Direct numerical simulations of multifluid flows in
  a vertical channel undergoing topology changes,'' \emph{Physical Review
  Fluids}, vol.~3, no.~8, p. 084401, 2018.

\bibitem[Chiodi(2020)]{chiodi2020advancement}
R.~M. Chiodi, \emph{Advancement of numerical methods for simulating primary
  atomization}.\hskip 1em plus 0.5em minus 0.4em\relax Cornell University,
  2020.

\bibitem[Jemison et~al.(2015)Jemison, Sussman, and
  Shashkov]{jemison2015filament}
M.~Jemison, M.~Sussman, and M.~Shashkov, ``Filament capturing with the
  multimaterial moment-of-fluid method,'' \emph{Journal of Computational
  Physics}, vol. 285, pp. 149--172, 2015.

\bibitem[Chirco et~al.(2022)Chirco, Maarek, Popinet, and
  Zaleski]{chirco2022manifold}
L.~Chirco, J.~Maarek, S.~Popinet, and S.~Zaleski, ``Manifold death: a volume of
  fluid implementation of controlled topological changes in thin sheets by the
  signature method,'' \emph{Journal of Computational Physics}, vol. 467, p.
  111468, 2022.

\bibitem[Saurabh et~al.(2021{\natexlab{a}})Saurabh, Gao, Fernando, Xu,
  Khanwale, Khara, Hsu, Krishnamurthy, Sundar, and
  Ganapathysubramanian]{saurabh2021industrial}
K.~Saurabh, B.~Gao, M.~Fernando, S.~Xu, M.~A. Khanwale, B.~Khara, M.-C. Hsu,
  A.~Krishnamurthy, H.~Sundar, and B.~Ganapathysubramanian, ``Industrial scale
  large eddy simulations with adaptive octree meshes using immersogeometric
  analysis,'' \emph{Computers \& Mathematics with Applications}, vol.~97, pp.
  28--44, 2021.

\bibitem[Saurabh et~al.(2021{\natexlab{b}})Saurabh, Ishii, Fernando, Gao, Tan,
  Hsu, Krishnamurthy, Sundar, and Ganapathysubramanian]{saurabh2021scalable}
K.~Saurabh, M.~Ishii, M.~Fernando, B.~Gao, K.~Tan, M.-C. Hsu, A.~Krishnamurthy,
  H.~Sundar, and B.~Ganapathysubramanian, ``Scalable adaptive pde solvers in
  arbitrary domains,'' in \emph{Proceedings of the International Conference for
  High Performance Computing, Networking, Storage and Analysis}, 2021, pp.
  1--15.

\bibitem[Fernando et~al.(2018)Fernando, Neilsen, Lim, Hirschmann, and
  Sundar]{fernando2018massively}
M.~Fernando, D.~Neilsen, H.~Lim, E.~Hirschmann, and H.~Sundar, ``Massively
  parallel simulations of binary black hole intermediate-mass-ratio
  inspirals,'' \emph{arXiv preprint arXiv:1807.06128}, 2018.

\bibitem[Popinet(2003)]{popinet2003gerris}
S.~Popinet, ``Gerris: a tree-based adaptive solver for the incompressible euler
  equations in complex geometries,'' \emph{Journal of computational physics},
  vol. 190, no.~2, pp. 572--600, 2003.

\bibitem[Bangerth et~al.(2012)Bangerth, Burstedde, Heister, and
  Kronbichler]{bangerth2012algorithms}
W.~Bangerth, C.~Burstedde, T.~Heister, and M.~Kronbichler, ``Algorithms and
  data structures for massively parallel generic adaptive finite element
  codes,'' \emph{ACM Transactions on Mathematical Software (TOMS)}, vol.~38,
  no.~2, pp. 1--28, 2012.

\bibitem[Khanwale et~al.(2023)Khanwale, Saurabh, Ishii, Sundar, Rossmanith, and
  Ganapathysubramanian]{Khanwale2023projection}
M.~A. Khanwale, K.~Saurabh, M.~Ishii, H.~Sundar, J.~A. Rossmanith, and
  B.~Ganapathysubramanian, ``A projection-based, semi-implicit time-stepping
  approach for the cahn-hilliard navier-stokes equations on adaptive octree
  meshes,'' \emph{Journal of Computational Physics}, vol. 475, p. 111874, 2023.

\bibitem[Guo et~al.(2017)Guo, Lin, Lowengrub, and Wise]{guo2017mass}
Z.~Guo, P.~Lin, J.~Lowengrub, and S.~M. Wise, ``Mass conservative and energy
  stable finite difference methods for the quasi-incompressible
  navier--stokes--cahn--hilliard system: Primitive variable and projection-type
  schemes,'' \emph{Computer Methods in Applied Mechanics and Engineering}, vol.
  326, pp. 144--174, 2017.

\bibitem[Shen and Yang(2015)]{shen2015decoupled}
J.~Shen and X.~Yang, ``Decoupled, energy stable schemes for phase-field models
  of two-phase incompressible flows,'' \emph{SIAM Journal on Numerical
  Analysis}, vol.~53, no.~1, pp. 279--296, 2015.

\bibitem[Anderson et~al.(1998)Anderson, McFadden, and
  Wheeler]{anderson1998diffuse}
D.~M. Anderson, G.~B. McFadden, and A.~A. Wheeler, ``Diffuse-interface methods
  in fluid mechanics,'' \emph{Annual review of fluid mechanics}, vol.~30,
  no.~1, pp. 139--165, 1998.

\bibitem[Feng(2006)]{feng2006fully}
X.~Feng, ``Fully discrete finite element approximations of the
  navier--stokes--cahn-hilliard diffuse interface model for two-phase fluid
  flows,'' \emph{SIAM journal on numerical analysis}, vol.~44, no.~3, pp.
  1049--1072, 2006.

\bibitem[Ishii et~al.(2019)Ishii, Fernando, Saurabh, Khara,
  Ganapathysubramanian, and Sundar]{ishii2019solving}
M.~Ishii, M.~Fernando, K.~Saurabh, B.~Khara, B.~Ganapathysubramanian, and
  H.~Sundar, ``Solving pdes in space-time: 4d tree-based adaptivity, mesh-free
  and matrix-free approaches,'' in \emph{Proceedings of the International
  Conference for High Performance Computing, Networking, Storage and Analysis},
  2019, pp. 1--61.

\bibitem[Fernando et~al.(2017)Fernando, Duplyakin, and
  Sundar]{fernando2017machine}
M.~Fernando, D.~Duplyakin, and H.~Sundar, ``Machine and application aware
  partitioning for adaptive mesh refinement applications,'' in
  \emph{Proceedings of the 26th International Symposium on High-Performance
  Parallel and Distributed Computing}, 2017, pp. 231--242.

\bibitem[Burstedde et~al.(2011)Burstedde, Wilcox, and
  Ghattas]{burstedde2011p4est}
C.~Burstedde, L.~C. Wilcox, and O.~Ghattas, ``p4est: Scalable algorithms for
  parallel adaptive mesh refinement on forests of octrees,'' \emph{SIAM Journal
  on Scientific Computing}, vol.~33, no.~3, pp. 1103--1133, 2011.

\bibitem[Sundar et~al.(2012)Sundar, Biros, Burstedde, Rudi, Ghattas, and
  Stadler]{sundar2012parallel}
H.~Sundar, G.~Biros, C.~Burstedde, J.~Rudi, O.~Ghattas, and G.~Stadler,
  ``Parallel geometric-algebraic multigrid on unstructured forests of
  octrees,'' in \emph{SC'12: Proceedings of the International Conference on
  High Performance Computing, Networking, Storage and Analysis}.\hskip 1em plus
  0.5em minus 0.4em\relax IEEE, 2012, pp. 1--11.

\bibitem[Sundar et~al.(2013)Sundar, Malhotra, and Biros]{sundar2013hyksort}
H.~Sundar, D.~Malhotra, and G.~Biros, ``{HykSort}: a new variant of hypercube
  quicksort on distributed memory architectures,'' in \emph{Proceedings of the
  27th international ACM conference on international conference on
  supercomputing}, 2013, pp. 293--302.

\bibitem[Hoefler et~al.(2010)Hoefler, Siebert, and
  Lumsdaine]{hoefler2010scalable}
T.~Hoefler, C.~Siebert, and A.~Lumsdaine, ``Scalable communication protocols
  for dynamic sparse data exchange,'' \emph{ACM Sigplan Notices}, vol.~45,
  no.~5, pp. 159--168, 2010.

\bibitem[Guo et~al.(2022)Guo, Cheng, Lin, Liu, and Lowengrub]{guo2022second}
Z.~Guo, Q.~Cheng, P.~Lin, C.~Liu, and J.~Lowengrub, ``Second order
  approximation for a quasi-incompressible navier-stokes cahn-hilliard system
  of two-phase flows with variable density,'' \emph{Journal of Computational
  Physics}, vol. 448, p. 110727, 2022.

\bibitem[Khanwale et~al.(2022)Khanwale, Saurabh, Ishii, Sundar, and
  Ganapathysubramanian]{dummy}
M.~A. Khanwale, K.~Saurabh, M.~Ishii, H.~Sundar, and B.~Ganapathysubramanian,
  ``Breakup dynamics in primary jet atomization using mesh-and
  interface-refined cahn-hilliard navier-stokes,'' \emph{arXiv e-prints}, pp.
  arXiv--2209, 2022.

\bibitem[Khanwale et~al.(2020)Khanwale, Lofquist, Sundar, Rossmanith, and
  Ganapathysubramanian]{khanwale2020simulating}
M.~A. Khanwale, A.~D. Lofquist, H.~Sundar, J.~A. Rossmanith, and
  B.~Ganapathysubramanian, ``Simulating two-phase flows with thermodynamically
  consistent energy stable cahn-hilliard navier-stokes equations on parallel
  adaptive octree based meshes,'' \emph{Journal of Computational Physics}, vol.
  419, p. 109674, 2020.

\bibitem[Tran et~al.(2022)Tran, Fernando, Saurabh, Ganapathysubramanian, Kirby,
  and Sundar]{tran2022scalable}
H.~D. Tran, M.~Fernando, K.~Saurabh, B.~Ganapathysubramanian, R.~M. Kirby, and
  H.~Sundar, ``A scalable adaptive-matrix spmv for heterogeneous
  architectures,'' in \emph{2022 IEEE International Parallel and Distributed
  Processing Symposium (IPDPS)}.\hskip 1em plus 0.5em minus 0.4em\relax IEEE,
  2022, pp. 13--24.

\bibitem[Tomov et~al.(2009)Tomov, Dongarra, Volkov, and Demmel]{tomov2009magma}
S.~Tomov, J.~Dongarra, V.~Volkov, and J.~Demmel, ``Magma library,'' \emph{Univ.
  of Tennessee and Univ. of California, Knoxville, TN, and Berkeley, CA}, 2009.

\bibitem[Heisler et~al.(2022)Heisler, Deshmukh, and Sundar]{heisler2022domain}
E.~Heisler, A.~Deshmukh, and H.~Sundar, ``: Domain specific language and code
  generation for finite element and finite volume in julia,'' in
  \emph{International Conference on Computational Science}.\hskip 1em plus
  0.5em minus 0.4em\relax Springer, 2022, pp. 118--132.

\bibitem[Aln{\ae}s et~al.(2015)Aln{\ae}s, Blechta, Hake, Johansson, Kehlet,
  Logg, Richardson, Ring, Rognes, and Wells]{alnaes2015fenics}
M.~Aln{\ae}s, J.~Blechta, J.~Hake, A.~Johansson, B.~Kehlet, A.~Logg,
  C.~Richardson, J.~Ring, M.~E. Rognes, and G.~N. Wells, ``The fenics project
  version 1.5,'' \emph{Archive of Numerical Software}, vol.~3, no. 100, 2015.

\end{thebibliography}

\end{document}